\documentclass[lettersize,journal]{IEEEtran}
\usepackage{algorithmic}
\usepackage{algorithm}
\usepackage{array}
\usepackage{textcomp}
\usepackage{stfloats}
\usepackage{url}
\usepackage{verbatim}
\usepackage{graphicx}
\usepackage{cite}
\usepackage{cmap}
\usepackage[T1]{fontenc}
\usepackage{amsmath,amssymb,amsfonts}
\usepackage{amsthm}
\usepackage{bm}
\usepackage{bbm}
\ifCLASSOPTIONcompsoc
	\usepackage[caption=false,font=normalsize,labelfont=sf,textfont=sf]{subfig}
\else
	\usepackage[caption=false,font=footnotesize]{subfig}
\fi
\newtheorem{theorem}{Theorem}
\newtheorem{lemma}{Lemma}

\newcommand*{\vertbar}{\rule[-1ex]{0.5pt}{2.5ex}}

\begin{document}

\title{Transmit Precoder Design Approaches for Dual-Function Radar-Communication Systems}

\author{Jacob Pritzker, James Ward\textsuperscript{1}\thanks{DISTRIBUTION STATEMENT A. Approved for public release. Distribution is unlimited. The work of Jacob Pritzker and James Ward was supported in part by the Office of the Undersecretary of Defense for Research and Engineering (OUSD(R\&E)) under Air Force Contract No. FA8702-15-D- 0001. Any opinions, findings, conclusions or recommendations expressed in this material are those of the author(s) and do not necessarily reflect the views of the OUSD(R\&E).},~\IEEEmembership{Fellow,~IEEE,} Yonina C. Eldar,~\IEEEmembership{Fellow,~IEEE}\thanks{J. Pritzker just finished a master's program at MIT (jacobpritzker@alum.mit.edu)}\thanks{J. Ward is an associate division head at MIT Lincoln Laboratory (jward@ll.mit.edu)}\thanks{Y. C. Eldar is a professor of electrical engineering at the Weizmann Institute of Science (yonina.eldar@weizmann.ac.il)}}



\maketitle

\begin{abstract}
As radio-frequency (RF) antenna, component and processing capabilities increase, the ability to perform multiple RF system functions from a common aperture is being realized. Conducting both radar and communications from the same system is potentially useful in vehicular, health monitoring, and surveillance settings. This paper considers multiple-input-multiple-output (MIMO) dual-function radar-communication (DFRC) systems in which the radar and communication modes use distinct baseband waveforms. A transmit precoder provides spatial multiplexing and power allocation among the radar and communication modes. Multiple precoder design approaches are introduced for a radar detection mode in which a total search volume is divided into dwells to be searched sequentially. The approaches are designed to enforce a reliance on radar waveforms for sensing purposes, yielding improved approximation of desired ambiguity functions over prior methods found in the literature. The methods are also shown via simulation to enable design flexibility, allowing for prioritization of either subsystem and specification of a desired level of radar or communication performance.
\end{abstract}

\section{Introduction}
\IEEEPARstart{I}{n} recent years, wireless communication systems have started using larger portions of the radio-frequency (RF) spectrum, some bands of which have previously been reserved for radar applications. In addition, both radar and communication systems benefit from utilizing larger bandwidths, which further congests the available spectrum. This has motivated the design of radar and communication systems which share information with one another in order to coordinate spectrum sharing, as well as dual-function radar-communication (DFRC) systems.

Numerous approaches to DFRC system design have been explored in the literature \cite{6Goverview,survey,survey2,survey3,survey4}. Some approaches have attempted to use a single waveform for sensing and communication purposes. For instance, \cite{preamble} utilizes the preamble of the 802.11ad communication protocol for sensing, while \cite{IEEE80211p} uses the IEEE 802.11p protocol for both sensing and communications. Finally, \cite{OFDM} analyzes the potential of using orthogonal frequency-division multiplexing (OFDM) communication waveforms for sensing. Other works have instead embedded data within legacy radar waveforms, such as by embedding data bits within continuous phase modulation of frequency-modulated constinuous-wave (FMCW) radar waveforms \cite{FMCW}, or within chirp waveforms \cite{chirp1, chirp2}. Alternative approaches considered designing new types of waveforms for dual-functional use \cite{JD1, JD2}, or utilizing index modulation, in which data are embedded in system parameters, such as antenna allocation and carrier frequency selection \cite{MAJoRCom, IM_FMCW, SPACOR, IM2, IM3}.

A different strategy is to utilize distinct radar and communications waveforms. Several approaches have imposed time \cite{time_division, time_division2} or frequency \cite{freqdiv, freqdiv2, freqdiv3, CRo_CRr} division in order to mitigate interference, which places an inherent tradeoff of system resources between radar and communication modes. Others have utilized transmit beamforming in the context of multiple-input-multiple-output (MIMO) DFRC systems \cite{ZF_beamforming, nullspace_beamforming, spatial_orthogonality, CRB, precoder_paper}. In such strategies, the spatial degrees of freedom are utilized to spatially multiplex waveforms towards communication receivers and areas to be searched.

The work \cite{CRB} develops beamformer design approaches which minimize the Cram\'er-Rao bound (CRB) on parameter estimates for a radar target while guaranteeing some desired communication signal-to-interference-plus-noise ratio (SINR) threshold for downlink receivers. The point target case considered is designed to track a target with an a priori angle estimate, while the extended target case has no directionality incorporated. The work \cite{precoder_paper} develops approaches for a similar scenario but allows for specification of a search dwell of interest. The methods introduced utilize a radar metric based on the mean-squared error (MSE) between the designed beampattern and a desired beampattern.

While these approaches are well-suited for target tracking, in which the goal is to acquire accurate parameter updates for known targets with prior parameter estimates, in our work, we develop approaches based on a new radar metric focused on a radar detection mode, in which the goal is initial target detection. We consider settings in which the total search volume is divided into dwells to be searched sequentially. This dwell specification, not possible using the methods from \cite{CRB}, aligns our setting with that of \cite{precoder_paper}.

Unlike the previous works \cite{CRB, precoder_paper}, which introduce methods guaranteeing some desired communication performance level, we introduce methods guaranteeing a user-prescribed level of radar performance, providing further design flexibility with the ability to prioritize either subsystem. We also introduce a method guaranteeing a user-prescribed level of communications performance, utilizing a new radar metric. In addition, our desired DFRC system is one that relies on good radar waveforms for sensing. We show via simulation below that all new methods exhibit improved approximation of desired ambiguity functions over prior methods in some scenarios of interest.

The rest of this paper is organized as follows. Section~\ref{section:models_metrics} introduces the system and signal models used, as well as the metrics considered to drive the precoder design methods. Section~\ref{section:guar} introduces the radar guarantee and communication guarantee design approaches and shows a variety of simulations to illustrate the functionality of these two new methods, as well as comparisons between them. Section~\ref{section:priority} introduces the radar priority approach, and Section~\ref{section:prior_methods} compares the new techniques to prior methods found in the literature. Finally, Section~\ref{section:conclusion} concludes the paper. Portions of this work appear in \cite{our_paper}, and much of this work appears in \cite{thesis}.

Throughout, we use the notation $(\cdot)^T$, $(\cdot)^*$, and $(\cdot)^H$ to denote transpose, conjugate, and Hermitian transpose, respectively. We represent vectors and matrices with bold lower and upper case letters, respectively. For a vector $\bm{v}$, we denote by $||\bm{v}||_2$ the $\ell_2$ norm of $\bm{v}$ and by $\text{diag}(\bm{v})$ the diagonal matrix whose diagonal elements are the elements of $\bm{v}$. We denote the trace of square matrix $\bm{A}$ by $\text{tr}(\bm{A})$. In addition, we use $\bm{0}_{a\times b}$ for the $a\times b$ 0-matrix and $\bm{I}_a$ for the $a\times a$ identity matrix. We use $\mathcal{S}_m^{+}$ to denote the set of $m\times m$ positive semidefinite matrices.

\section{Models and Metrics}
\label{section:models_metrics}
\noindent In this section, we introduce the system and signal models, as well as the metrics used to drive precoder design.
\subsection{System and Signal Model}
\begin{figure}[!t]
\centering
\includegraphics[width=8cm]{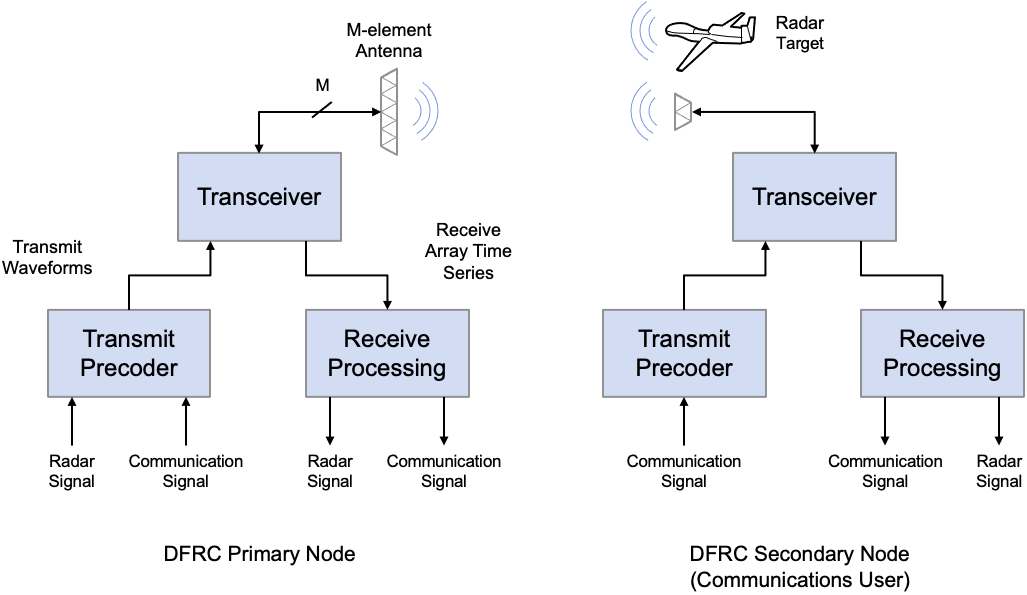}
\caption{DFRC system model.}
\label{fig:DFRC}
\end{figure}

We consider a DFRC system similar to that in \cite{precoder_paper} comprised of two types of nodes, termed primary and secondary, as shown in Fig.~\ref{fig:DFRC}.

The primary node provides monostatic radar surveillance, including both radar waveform transmission and radar receive signal processing, as well as a communications service to secondary nodes. A secondary node is a communications user that receives the transmissions from the primary node. A secondary node may also transmit signals intended for the primary node, which will perform communications receive processing to decode these transmissions. The primary node has an $M$-element array antenna, as well as a transceiver at each element to handle up- and down-conversion, conversion between analog and digital, and necessary RF filtering and amplification. The signals into and out of the transceiver at each element are discrete-time complex baseband signals. The transmit precoder takes as input the radar and communications waveforms and produces the baseband signals that, after transceiver filtering, power amplification, and upconversion, are transmitted from the $M$ array elements.

We denote by $K_c$ the number of secondary nodes and by $K_r$ the number of transmitted radar waveforms. We collect the transmitted baseband communications and radar waveforms in the $K_c\times1$ vector $\bm{c}[n]$ and the $K_r\times1$ vector $\bm{r}[n]$, respectively, and let $\bm{s}[n] = \big{[}\bm{c}^T[n], \bm{r}^T[n]\big{]}^T$. To facilitate precoder development, we view each of the sequences in $\bm{s}[n]$ as a wide-sense stationary (WSS) random process, with zero mean and unit power, and we assume they are uncorrelated with one another, meaning
\begin{equation}
\label{eq:uncorrelated}
\mathbb{E}\Big{[}\bm{s}[n]\bm{s}^H[n]\Big{]} = \bm{I}_{K_c+K_r}.
\end{equation}
We represent the precoder as an $M\times(K_c+K_r)$ matrix $\bm{W}=[\bm{W_c},\bm{W_r}]$ containing transmit directionality and fractional power allocations for the different waveforms collected in $\bm{s}[n]$.

For instance, to transmit a waveform in some direction, we could choose the associated column of $\bm{W}$ to be a scaled version of the transmit steering vector in that direction. For a uniform linear array with inter-element spacing $d$, we define the transmit steering vector in direction $\theta$ as
\begin{equation}
\label{eq:steering_vec}
\bm{a}(\theta) \triangleq \begin{bmatrix}
1 & e^{j\frac{2\pi d}{\lambda}\sin\theta} & \ldots & e^{j(M-1)\frac{2\pi d}{\lambda}\sin\theta}
\end{bmatrix}^T,
\end{equation}
where $\lambda$ is the carrier wavelength. We strive to design more sophisticated precoders than simple steering vectors.

The transmitted baseband waveforms from the DFRC primary node, after precoding, are represented as the $M\times1$ vector
\begin{equation}
\bm{x}[n] = \sqrt{P_e}\bm{W}\bm{s}[n] = \sqrt{P_e}(\bm{W_c}\bm{c}[n] + \bm{W_r}\bm{r}[n]),
\end{equation}
where $P_e$ denotes the transmit power per element, and $\bm{W_c}$ and $\bm{W_r}$ represent the communications and radar components of the precoder, respectively. We observe that the precoder defines the transmit spatial covariance matrix
\begin{equation}
\bm{R_{xx}} \triangleq \mathbb{E}\big{[}\bm{x}[n]\bm{x}^H[n]\big{]} = P_e\bm{W}\bm{W}^H,
\end{equation}
where we used \eqref{eq:uncorrelated}. In our precoder design approaches, we will enforce a per-antenna power constraint that the diagonal elements of $\bm{R_{xx}}$ be equal to $P_e$, or equivalently, that the diagonal elements of $\bm{W}\bm{W}^H$ all equal 1.

The complex baseband signal incident upon a target or secondary node is
\begin{equation}
\label{eq:incident}
y[n] = \sqrt{P_e}\bm{h}^H\bm{W}\bm{s}[n-m],
\end{equation}
where $\bm{h}$ is the vector of channels to this point from the elements of the DFRC primary node, and $m$ is the delay. For the $k$-th secondary node, we specialize our notation to $y[n]=y_k[n]$, $\bm{h}=\bm{h_k}$, and $m=n_k$. In addition, we assume additive white noise of variance $\sigma_k^2$ uncorrelated with $\bm{s}[m]$ for all $m$ at the secondary node, which we denote by $q_k[n]$. Thus, the signal observed at the $k$-th secondary node is
\begin{equation}
\label{eq:comm_rec}
y_k[n] = \sqrt{P_e}\bm{h_k}^H\bm{W}\bm{s}[n-n_k] + q_k[n].
\end{equation}
We collect the channels from DFRC primary node to secondary nodes, which we assume to be known, in the $K_c\times M$ complex matrix
\begin{equation*}
\bm{H_c} =
\begin{bmatrix}
\vertbar & \vertbar & & \vertbar \\
\bm{h_1} & \bm{h_2} & \cdots & \bm{h_{K_c}} \\
\vertbar & \vertbar & & \vertbar
\end{bmatrix}^H.
\end{equation*}

If there is a line of sight path to the $k$-th secondary node, and it is at angle $\theta_k$ relative to the primary node, then $\bm{h_k}$ is a scaled version of $\bm{a}(\theta_k)$. We then denote the observed signal at the $k$-th secondary node by
\begin{equation}
\label{eq:comm_rec_LoS}
y_k[n] = \alpha_k\bm{a}^H(\theta_k)\bm{W}\bm{s}[n-n_k] + q_k[n],
\end{equation}
where $\alpha_k$ is a complex amplitude determined by $P_e$ and the one-way propagation loss. In this case, we define the secondary node input signal-to-noise ratio (SNR) to be $\xi_{ic}^k=|\alpha_k|^2/\sigma_k^2$, which has components due to the desired communications transmission and undesired components due to the radar waveform transmissions and possibly other secondary node communications transmissions.

Extending \eqref{eq:incident} for the radar subsystem, the reflected waveform received at the primary node due to a target with channel $\bm{h}$ and two-way propagation delay $\ell$ is
\begin{equation}
\label{eq:primary_rec}
\bm{y_p}[n] = \sqrt{P_e}\bm{h}^{*}\bm{h}^H\bm{W}\bm{s}[n-\ell].
\end{equation}
We also assume there to be additive noise at each element of the primary node antenna array, which we collect in $M\times1$ vector $\bm{q_p}[n]$. We assume that this noise is spatially and temporally white, has variance $\sigma_p^2$, and is uncorrelated with $\bm{s}[m]$ for all $m$.

If the radar channel is a single line-of-sight path to and from this target, and it is at angle $\theta_r$, then the observations at the primary node are given by
\begin{equation}
\label{eq:primary_rec_los}
\bm{y_p}[n] = \alpha_r\bm{a}^{*}(\theta_r)\bm{a}^H(\theta_r)\bm{W}\bm{s}[n-\ell] + \bm{q_p}[n],
\end{equation}
where $\alpha_r$ denotes a complex amplitude determined by $P_e$, the two-way propagation loss, and the target's radar cross section. We define the target input SNR to be $\xi_{ir}=|\alpha_r|^2/\sigma_p^2$, which is due to a single element on transmit and receive, and contains contributions from both the radar and communications waveform transmissions.

\subsection{Radar and Communications Metrics}
Our interest is in designing the precoder matrices $\bm{W_c}$ and $\bm{W_r}$ to provide both primary node radar surveillance performance and primary-to-secondary node communications performance. As such, we strive to transmit good radar waveforms with high power in directions of interest, and to transmit communications waveforms towards secondary nodes while minimizing interference at these secondary nodes. We consider as a radar surveillance metric the output SINR of a radar receiver that does receive beamforming and matched filter radar waveform processing, though it is recognized that the actual receiver processing will likely differ from simple beamforming and matched filtering.

To develop our metric, we rewrite \eqref{eq:primary_rec_los} with the radar and communication waveforms separated as follows:
\begin{equation}
\label{eq:radar_SINR_dev1}
\bm{y_p}[n] = \alpha_r\bm{a}^{*}(\theta_r)\bm{a}^H(\theta_r)(\bm{W_r}\bm{r}[n-\ell] + \bm{W_c}\bm{c}[n-\ell]) + \bm{q_p}[n].
\end{equation}
Next, we consider the output of applying receive beamformer $\bm{a}^T(\theta_r)$:
\begin{alignat}{1}
\label{eq:radar_SINR_dev2}
z_p[n] & \triangleq \bm{a}^T(\theta_r)\bm{y_p}[n] \\
\nonumber
& = \alpha_rM\bm{a}^H(\theta_r)\bm{W_r}\bm{r}[n-\ell] + \alpha_rM\bm{a}^H(\theta_r)\bm{W_c}\bm{c}[n-\ell] \\
\label{eq:radar_SINR_dev5}
& \quad + \bm{a}^T(\theta_r)\bm{q_p}[n] \\
\label{eq:radar_SINR_dev6}
& \triangleq z_{p,r}[n] + z_{p,c}[n] + z_{p,n}[n],
\end{alignat}
where we used the fact that $\bm{a}^T(\theta_r)\bm{a}^*(\theta_r)=M$. In the final expression, we separated out the output of the beamformer into the sum of a radar term, a communications term, and a noise term. In developing our radar metric, we treat the communications transmissions as interference. While these communications waveforms are known at the primary node and can be used in receive processing, we structure the metric in this way to enforce reliance on radar transmissions for sensing, thus better approximating desired ambiguity functions.

The sample-level radar SINR for a target at angle $\theta_r$, after beamforming, is
\begin{equation}
\xi_r \triangleq \frac{\mathbb{E}\Big{[}\big{|}z_{p,r}[n]\big{|}^2\Big{]}}{\mathbb{E}\Big{[}\big{|}z_{p,c}[n]\big{|}^2\Big{]}+\mathbb{E}\Big{[}\big{|}z_{p,n}[n]\big{|}^2\Big{]}}.
\end{equation}
We next compute the expected squared magnitude of each term:
\begin{equation}
\label{eq:radar_SINR_radar1}
\mathbb{E}\Big{[}\big{|}z_{p,r}[n]\big{|}^2\Big{]} = |\alpha_r|^2M^2\bm{a}^H(\theta_r)\bm{W_r}\bm{W_r}^H\bm{a}(\theta_r),
\end{equation}
where we used \eqref{eq:uncorrelated}. Similarly, we have
\begin{equation}
\label{eq:radar_SINR_comm1}
\mathbb{E}\Big{[}\big{|}z_{p,c}[n]\big{|}^2\Big{]} = |\alpha_r|^2M^2\bm{a}^H(\theta_r)\bm{W_c}\bm{W_c}^H\bm{a}(\theta_r).
\end{equation}
Finally,
\begin{equation}
\label{eq:radar_SINR_noise1}
\mathbb{E}\Big{[}\big{|}z_{p,n}[n]\big{|}^2\Big{]} = \mathbb{E}\Big{[}\big{|}\bm{a}^T(\theta_r)\bm{q_p}[n]\big{|}^2\Big{]} = \sigma_p^2M,
\end{equation}
where we used our assumption that the noise is spatially and temporally white with variance $\sigma_p^2$. Thus,
\begin{equation}
\xi_r = \frac{M\xi_{ir}\bm{a}^H(\theta_r)\bm{W_r}\bm{W_r}^H\bm{a}(\theta_r)}{M\xi_{ir}\bm{a}^H(\theta_r)\bm{W_c}\bm{W_c}^H\bm{a}(\theta_r) + 1},
\end{equation}
where we use the input SNR $\xi_{ir}=|\alpha_r|^2/\sigma_p^2$.

For purposes of precoder design, the target SINR at the output of a matched filter will be used as our radar metric. Treating the communications interference term as white noise for this purpose, the processing gain of a matched filter will be the radar pulse length $N_r$ (assuming constant modulus radar waveforms), yielding radar SINR
\begin{equation}
\label{eq:radar_SINR}
\xi_{or}(\bm{W_c}, \bm{W_r}, \theta_r, \xi_{ir}) = \frac{N_rM\xi_{ir}\bm{a}^H(\theta_r)\bm{W_r}\bm{W_r}^H\bm{a}(\theta_r)}{M\xi_{ir}\bm{a}^H(\theta_r)\bm{W_c}\bm{W_c}^H\bm{a}(\theta_r) + 1}.
\end{equation}
The input $\xi_{ir}$ will be a user-prescribed input in precoder design.

For communications, we assume a pulse-amplitude-modulation (PAM) scheme in which symbols from some constellation digitally modulate some pulse shape. Our communications metric is the secondary node communication receiver output SINR after symbol pulse matched filtering, which in receiver processing would precede a symbol decoding step. We once again recognize that different receiver processing may be employed in practice.

To develop this metric, we rewrite \eqref{eq:comm_rec} with the communications transmission intended for the $k$-th secondary node separated out, as follows:
\begin{alignat}{1}
\nonumber
y_k[n] &= \sqrt{P_e}\bm{h_k}^H\bm{w_k}\bm{c}_k[n-n_k] \\
& \quad + \sqrt{P_e}\bm{h_k}^H\sum_{j\neq k}\bm{w_j}\bm{s}_j[n-n_k] + q_k[n] \\
&\triangleq y_{k,k}[n] + y_{k,i}[n] + q_k[n].
\end{alignat}
Here, $\bm{w_j}$ denotes the $j$-th column of $\bm{W}$, $\bm{c}_k[n]$ denotes the $k$-th sequence in $\bm{c}[n]$, and $\bm{s}_j[n]$ denotes the $j$-th sequence in $\bm{s}[n]$. In addition, $y_{k,k}[n]$ is the component of the observations intended for the $k$-th secondary node, and $y_{k,i}[n]$ denotes interference. Our sample-level SINR is
\begin{equation}
\xi_c \triangleq \frac{\mathbb{E}\Big{[}\big{|}y_{k,k}[n]\big{|}^2\Big{]}}{\mathbb{E}\Big{[}\big{|}y_{k,i}[n]\big{|}^2\Big{]}+\mathbb{E}\Big{[}\big{|}q_k[n]\big{|}^2\Big{]}}.
\end{equation}

We compute the expected squared magnitude terms as follows:
\begin{equation}
\mathbb{E}\Big{[}\big{|}y_{k,k}[n]\big{|}^2\Big{]} = \mathbb{E}\Big{[}\big{|}\sqrt{P_e}\bm{h_k}^H\bm{w_k}\bm{c}_k[n-n_k]\big{|}^2\Big{]} = P_e|\bm{h_k}^H\bm{w_k}|^2,
\end{equation}
where we used \eqref{eq:uncorrelated}. Next,
\begin{alignat}{1}
\label{eq:comm_SINR_int_dev1}
\mathbb{E}\Big{[}\big{|}y_{k,i}[n]\big{|}^2\Big{]} &= \mathbb{E}\Bigg{[}\bigg{|}\sqrt{P_e}\bm{h_k}^H\sum_{j\neq k}\bm{w_j}\bm{s}_j[n-n_k]\bigg{|}^2\Bigg{]} \\
\label{eq:comm_SINR_int_dev3}
&= P_e\sum_{j\neq k}|\bm{h_k}^H\bm{w_j}|^2,
\end{alignat}
where we again used \eqref{eq:uncorrelated}. Finally, $\mathbb{E}\Big{[}\big{|}q_k[n]\big{|}^2\Big{]} = \sigma_k^2.$ Thus,
\begin{equation}
\xi_c = \frac{P_e|\bm{h_k}^H\bm{w_k}|^2}{P_e\sum_{j\neq k}|\bm{h_k}^H\bm{w_j}|^2 + \sigma_k^2}.
\end{equation}

As with the radar SINR, we approximate the receiver processing gain of applying a matched filter to the communications pulse shape by the length of the pulse. A symbol length of $N_c^k$ samples is assumed for the $k$-th secondary node, so that $\lfloor N_r/N_c^k\rfloor$ symbols are transmitted with each radar pulse. The output SINR is then
\begin{equation}
\label{eq:comm_SINR}
\xi_{oc}(\bm{W_c}, \bm{W_r}, k) = \frac{N_c^k|\bm{h_k}^H\bm{w_k}|^2}{\sum_{j\neq k}|\bm{h_k}^H\bm{w_j}|^2 + \sigma_k^2/P_e}.
\end{equation}
We emphasize that the radar transmission is treated as interference at the communication reciever.

\subsection{Beampatterns}
As a useful assessment of precoders, we define different types of beampatterns as normalized directional transmit gains. For the $j$-th column of $\bm{W}$, we define the beampattern associated with that column as
\begin{equation}
\label{eq:bp_single}
\text{B}^{(j)}(\theta) \triangleq \frac{\mathbb{E}\bigg{[}\Big{|}\bm{a}^H(\theta)\Big{(}\sqrt{P_e}\bm{w_j}\Big{)}\bm{s}_j[n]\Big{|}^2\bigg{]}}{\text{tr}\bigg{(}\Big{(}\sqrt{P_e}\bm{w_j}\Big{)}\Big{(}\sqrt{P_e}\bm{w_j}\Big{)}^H\bigg{)}}= \frac{|\bm{a}^H(\theta)\bm{w_j}|^2}{||\bm{w_j}||_2^2},
\end{equation}
where we used \eqref{eq:uncorrelated}. In the middle expression, the numerator term is the expected power from the $j$-th waveform transmitted to a point in direction $\theta$. The denominator term is the power across the $M$ antenna elements from this waveform.

We can similarly define beampatterns associated with the communication and radar components of the precoder, as well as that associated with the total precoder. Specifically, the communication beampattern is
\begin{equation}
\label{eq:bp_c}
\text{B}_c(\theta) \triangleq \frac{||\bm{a}^H(\theta)\bm{W_c}||_2^2}{\text{tr}(\bm{W_c}\bm{W_c}^H)},
\end{equation}
the radar beampattern is
\begin{equation}
\label{eq:bp_r}
\text{B}_r(\theta) \triangleq \frac{||\bm{a}^H(\theta)\bm{W_r}||_2^2}{\text{tr}(\bm{W_r}\bm{W_r}^H)},
\end{equation}
and the total beampattern is
\begin{equation}
\label{eq:bp}
\text{B}(\theta) \triangleq \frac{||\bm{a}^H(\theta)\bm{W}||_2^2}{\text{tr}(\bm{W}\bm{W}^H)}.
\end{equation}
Since
\begin{equation}
||\bm{a}^H(\theta)\bm{W}||_2^2 = ||\bm{a}^H(\theta)\bm{W_c}||_2^2 + ||\bm{a}^H(\theta)\bm{W_r}||_2^2,
\end{equation}
we see that
\begin{equation}
\text{B}(\theta) = \frac{\text{tr}(\bm{W_c}\bm{W_c}^H)\text{B}_c(\theta) + \text{tr}(\bm{W_r}\bm{W_r}^H)\text{B}_r(\theta)}{\text{tr}(\bm{W}\bm{W}^H)}.
\end{equation}
Therefore, the total beampattern is a weighted combination of the communication and radar beampatterns, with the weighting determined by the relative power allotted to each subsystem.

We plot these different types of beampatterns for different precoders below in order to gain insight into how design choices affect precoder design. We strive to design good beampatterns particular to the two modes of our system.

\section{Radar and Communication Guarantee Approaches}
\label{section:guar}
\noindent To design our precoder $\bm{W}$, we consider optimization problems with different constraints and objectives. Our first two methods, described in this section, allow for a user-prescribed radar or communication performance level, with remaining system resources being allotted to the other subsystem.

\subsection{Radar Guarantee}
Our first precoder design method is termed {\em radar guarantee}. For this method, we enforce a specified radar SINR $\Gamma_r$ performance for a worst-case input target SNR $\xi_{ir}$ over a discrete search sector $\Theta$ as a constraint. We note that this $\Gamma_r$ specification must be within the resource constraints of the DFRC system. Any system power resources remaining are utilized for communicating with the secondary nodes. The approach we take for the communications subsystem is to maximize the minimum communication SINR across all secondary nodes to best service them all. In addition, an equal power per antenna constraint is enforced

In order to design a precoder with these constraints and objective, we aim to solve the following problem:
\begin{subequations}
\label{eq:orig_rguar}
\begin{alignat}{2}
\label{eq:orig_rguar1}
&\max_{\bm{W_c}, \bm{W_r}, \Gamma_c} &&\Gamma_c \\
\label{eq:orig_rguar2}
&\text{subject to } \quad && \bm{R}=\bm{W_c}\bm{W_c}^H+\bm{W_r}\bm{W_r}^H, \\
\label{eq:orig_rguar3}
& && [\bm{R}]_{m,m} = 1, \quad m=1, \ldots, M, \\
\label{eq:orig_rguar4}
& && \xi_{oc}(\bm{W_c},\bm{W_r},k) \ge \Gamma_c, \quad k=1, \ldots, K_c, \\
\label{eq:orig_rguar5}
& && \xi_{or}(\bm{W_c},\bm{W_r},\theta,\xi_{ir}) \ge \Gamma_r, \quad \forall \theta \in \Theta,
\end{alignat}
\end{subequations}
where $\xi_{or}$ and $\xi_{oc}$ are defined by \eqref{eq:radar_SINR} and \eqref{eq:comm_SINR}, respectively. We note that by inspection of \eqref{eq:radar_SINR}, the radar SINR increases as the input SNR increases. Therefore, \eqref{eq:orig_rguar5} guarantees a radar SINR of at least $\Gamma_r$ for a target at any angle $\theta\in\Theta$ with any input SNR greater than or equal to $\xi_{ir}$.

Problem \eqref{eq:orig_rguar} is not convex. To relax the problem to a convex one, we recast it in terms of transmit covariance matrices $\bm{R} \triangleq \bm{W}\bm{W}^H$ and $\bm{R_k} \triangleq \bm{w_k}\bm{w_k}^H$ for $k=1,\ldots,K_c$. With this notation:
\begin{alignat}{1}
\nonumber
\xi_{oc}(\bm{W_c}, \bm{W_r}, k) &= \frac{N_c^k\bm{h_k}^H\bm{R_k}\bm{h_k}}{\bm{h_k}^H\bm{R}\bm{h_k}-\bm{h_k}^H\bm{R_k}\bm{h_k}+\sigma_k^2/P_e} \\
\label{eq:R_xi_oc}
& \triangleq \xi_{oc}'(\bm{R},\bm{R_k},k)
\end{alignat}
and
\begin{alignat}{1}
\nonumber
\xi_{or}(\bm{W_c}, \bm{W_r}, \theta, \xi_{ir}) &= \frac{N_rM\xi_{ir}\bm{a}^H(\theta)(\bm{R}-\sum_{k=1}^{K_c}\bm{R_k})\bm{a}(\theta)}{M\xi_{ir}\bm{a}^H(\theta)(\sum_{k=1}^{K_c}\bm{R_k})\bm{a}(\theta) + 1}\\
\label{eq:R_xi_or}
&\triangleq \xi_{or}'\bigg{(}\bm{R},\sum_{k=1}^{K_c}\bm{R_k},\theta,\xi_{ir}\bigg{)}.
\end{alignat}
For a fixed $\Gamma_c$, the following feasibility problem is convex as it has only linear and semidefinite constraints:
\begin{subequations}
\label{eq:guar}
\begin{alignat}{2}
\label{eq:guar1}
&\text{Find } && \bm{R}, \{\bm{R_k}\}_{k=1}^{K_c} \\
\label{eq:guar2}
&\text{subject to } \quad && \bm{R} \in \mathcal{S}_M^{+}, \\
\label{eq:guar3}
& && \bm{R_k} \in \mathcal{S}_M^{+}, \quad k=1, \ldots, K_c, \\
\label{eq:guar4}
& && \bm{R} - \sum_{k=1}^{K_c} \bm{R_k} \in \mathcal{S}_M^{+}, \\
\label{eq:guar5}
& && [\bm{R}]_{m,m} = 1, \quad m=1, \ldots, M, \\
\label{eq:guar6}
& && \xi_{oc}'(\bm{R},\bm{R_k},k) \ge \Gamma_c, \quad k=1, \ldots, K_c, \\
\label{eq:guar7}
& && \xi_{or}'\bigg{(}\bm{R},\sum_{k=1}^{K_c}\bm{R_k},\theta,\xi_{ir}\bigg{)} \ge \Gamma_r, \quad \forall \theta \in \Theta.
\end{alignat}
\end{subequations}
We assume throughout that $\Gamma_r$, $\Gamma_c$, and all noise variances are strictly positive. This problem enables us to find precoders satisfying a specified radar SINR and a specified communication SINR constraint, as stated in the following theorem:

\begin{theorem}
\label{thm:cov_iff}
The problem \eqref{eq:guar} is feasible if and only if there exist precoders $\bm{W_c}$ ($M\times K_c$) and $\bm{W_r}$ ($M\times M$) satisfying the per-antenna power constraint, achieving communication SINRs all at least $\Gamma_c$, and achieving radar SINRs all at least $\Gamma_r$ for worst-case input SNR $\xi_{ir}$ across search sector $\Theta$. Furthermore, when \eqref{eq:guar} is feasible, if we denote some feasible solution via $\widetilde{\bm{R}}$, $\{\widetilde{\bm{R_k}}\}_{k=1}^{K_c}$, then by defining columns of $\bm{W_c}$ via
\begin{equation}
\label{eq:cov_iff_wk}
\bm{w_k} = \Big{(}\bm{h_k}^H\widetilde{\bm{R_k}}\bm{h_k}\Big{)}^{-1/2}\widetilde{\bm{R_k}}\bm{h_k},\quad k=1,\ldots,K_c
\end{equation}
and choosing $\bm{W_r}$ such that
\begin{equation}
\label{eq:cov_iff_Wr}
\bm{W_r}\bm{W_r}^H = \widetilde{\bm{R}}-\bm{W_c}\bm{W_c}^H,
\end{equation}
we construct precoders satisfying the constraints \eqref{eq:orig_rguar} for fixed $\Gamma_c$.
\end{theorem}
\begin{IEEEproof}
See Appendix~\ref{app:cov_iff}, which is based on a similar derivation in \cite{precoder_paper}.
\end{IEEEproof}

Using the above formulation, we perform a bisection search over possible values for $\Gamma_c$. The final version of the radar guarantee approach is:
\begin{enumerate}
\item Determine a range of potential values for $\Gamma_c$, the smallest $\Gamma_{c,min}$ being some chosen value, and the largest $\Gamma_{c,max}$ determined from available power. By inspection of \eqref{eq:R_xi_oc}, the largest possible output SINR for the $k$-th secondary node is $P_eN_c^k\bm{h_k}^H\bm{R}\bm{h_k}/\sigma_k^2$, achieved if all power is put into the $k$-th column of $\bm{W}$. Since $\bm{R}$ is positive semidefinite and constrained to have trace $M$, its largest possible eigenvalue is $M$, meaning $P_eN_c^k\bm{h_k}^H\bm{R}\bm{h_k}/\sigma_k^2 \le P_eMN_c^k\bm{h_k}^H\bm{h_k}/\sigma_k^2$. Minimizing this across all secondary nodes sets our value of $\Gamma_{c,max}$.
\item Perform a bisection search, fixing values of $\Gamma_c$ and for each attempting to solve the following problem:
\begin{subequations}
\label{eq:rguar}
\begin{alignat}{2}
\label{eq:rguar1}
&\max_{\bm{R}, \{\bm{R_k}\}_{k=1}^{K_c}, t} &&t \\
\label{eq:rguar2}
&\text{subject to } \quad && \bm{R} \in \mathcal{S}_M^{+}, \\
\label{eq:rguar3}
& && \bm{R_k} \in \mathcal{S}_M^{+}, \quad k=1, \ldots, K_c, \\
\label{eq:rguar4}
& && \bm{R} - \sum_{k=1}^{K_c} \bm{R_k} \in \mathcal{S}_M^{+}, \\
\label{eq:rguar5}
& && [\bm{R}]_{m,m} = 1, \quad m=1, \ldots, M, \\
\label{eq:rguar6}
& && \xi_{oc}'(\bm{R},\bm{R_k},k) \ge \Gamma_c, \quad k=1, \ldots, K_c, \\
\label{eq:rguar7}
& && \xi_{or}'\bigg{(}\bm{R},\sum_{k=1}^{K_c} \bm{R_k},\theta,\xi_{ir}\bigg{)} \ge \Gamma_r, \quad \forall \theta \in \Theta, \\
\nonumber
& && \frac{P_e}{\sigma_k^2}\bm{h_k}^H\Big{(}(N_c^k+\Gamma_c)\bm{R_k}-\Gamma_c\bm{R}\Big{)}\bm{h_k} \\
\label{eq:rguar8}
& && \quad - \Gamma_c \ge t, \quad k=1, \ldots, K_c.
\end{alignat}
\end{subequations}
Constraint \eqref{eq:rguar8}, which derives from a rearrangement of \eqref{eq:R_xi_oc}, serves to push us closer to the globally optimal value for $\Gamma_c$.
\item Start the search with $\Gamma_c = \Gamma_{c,min}$. If \eqref{eq:rguar} is feasible, continue, and in successive steps, choose $\Gamma_c$ as the midpoint of the remaining search range. Otherwise, terminate.
\item At each search step, if \eqref{eq:rguar} is feasible, set the used value of $\Gamma_c$ as the new lower limit, and if it is infeasible, as the new upper limit. Stop once the search has narrowed to a range of at most a user-specified $\epsilon_c$.
\item Denoting the optimal solution to \eqref{eq:rguar} with $\Gamma_c$ set to be the lower limit on the final search range by $\widetilde{\bm{R}}$, $\{\widetilde{\bm{R_k}}\}_{k=1}^{K_c}$, $\widetilde{t}$, form the columns of $\bm{W_c}$ as
\begin{equation}
\label{eq:rguar_wk}
\bm{w_k} = \Big{(}\bm{h_k}^H\widetilde{\bm{R_k}}\bm{h_k}\Big{)}^{-1/2}\widetilde{\bm{R_k}}\bm{h_k},\quad k=1,\ldots,K_c.
\end{equation}
\item Choose $\bm{W_r}$ such that
\begin{equation}
\label{eq:rguar_Wr}
\bm{W_r}\bm{W_r}^H = \widetilde{\bm{R}}-\bm{W_c}\bm{W_c}^H.
\end{equation}
We perform this via eigen-decomposition.
\end{enumerate}
Since \eqref{eq:rguar} and \eqref{eq:guar} share feasible regions over $\bm{R}$, $\{\bm{R_k}\}_{k=1}^{K_c}$, our bisection search approach yields final precoders achieving minimum communication SINR within $\epsilon_c$ of the optimal value to \eqref{eq:orig_rguar} by Theorem~\ref{thm:cov_iff}.

To see this method's functionality, we analyze a simple example. Here and in all examples in this paper, we use an $M=10$-element uniform linear array with half-wavelength spacing at the primary node, and we do our bisection search in log-space with $\epsilon_c=0.1$ dB. The sector $\Theta$ we specify is half of a beamwidth centered at broadside, discretized in increments of one-tenth of a beamwidth in sine space. The radar guarantee is 15 dB radar SINR for a target with worst-case input SNR $-34$ dB, using a length-100 radar pulse. We place a single secondary node with $-5$ dB input SNR at 17\textdegree, which is right near the peak of the first sidelobe of the beampattern of a length-10 uniformly weighted beamformer, and we assume a line-of-sight channel to this secondary node. We use a length-10 communication symbol pulse shape. To solve \eqref{eq:rguar}, as well as subsequent optimization problems, we use the toolbox CVX, a package for specifying and solving convex programs \cite{CVX1, CVX2}.

\begin{figure}[!t]
\centering
\subfloat{\includegraphics[width=4.5cm]{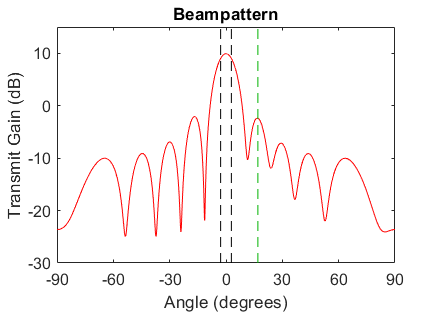}}
\subfloat{\includegraphics[width=4.5cm]{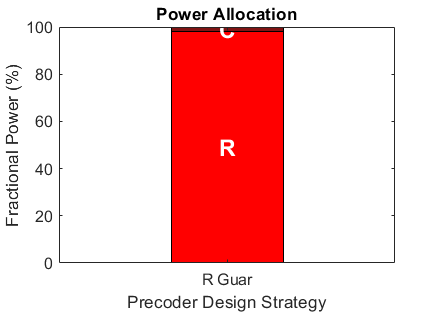}}
\hfil
\subfloat{\includegraphics[width=4.5cm]{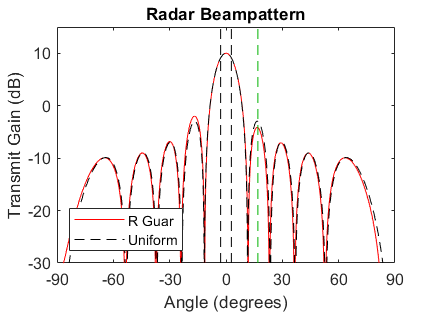}}
\subfloat{\includegraphics[width=4.5cm]{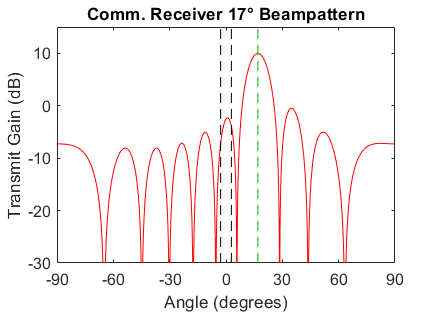}}
\caption{Radar guarantee with minimal excess power.}
\label{fig:rguar1}
\end{figure}

We could gain up to $N_rM^3$, which here is 50 dB, of radar SINR via receive processing. This can be seen by inspection of \eqref{eq:R_xi_or}. Since $\bm{R}$ is positive semidefinite and has trace $M$, its largest possible eigenvalue is $M$, meaning $\bm{a}^H(\theta)\bm{R}\bm{a}(\theta)\le M^2$ for any $\theta$ since $\bm{a}^H(\theta)\bm{a}(\theta)=M$. Since each $\bm{R_k}$ is positive semidefinite, $\bm{a}^H(\theta)\bm{R_k}\bm{a}(\theta)\ge0$. This means the numerator is at most $N_rM^3\xi_{ir}$, and the denominator is at least 1, meaning the maximum processing gain is $N_rM^3$. Since we need to achieve 15 dB SINR for a target with input SNR $-34$ dB, there is minimal excess power and flexibility. The results are shown in Fig.~\ref{fig:rguar1}.

The power allocation plot shows the fraction allotted to the radar subsystem on bottom and that allotted to communications on top. We see that nearly all power (roughly 97.9\%) is allocated to the radar subsystem. The low fractional power allocated for communications, coupled with the relative transmit gains of radar and communications within the search sector, causes the communications waveform contribution to target scattered power to be much less than that of the radar waveform. In the radar beampattern plot, we also plot the beampattern of a uniform beamformer for comparison. These curves are very similar, with the radar guarantee radar beampattern having slightly lower transmit gain towards the secondary node, as it is trying to maximize communication SINR at the secondary node. Radar guarantee is able to achieve roughly 4.7 dB of communication SINR.

\begin{figure}[!t]
\centering
\subfloat{\includegraphics[width=4.5cm]{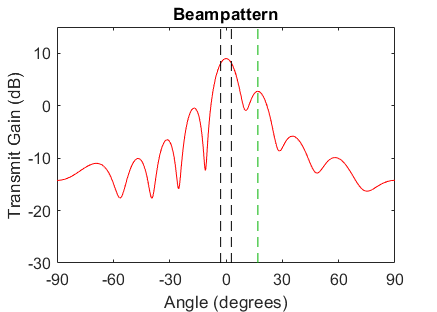}}
\subfloat{\includegraphics[width=4.5cm]{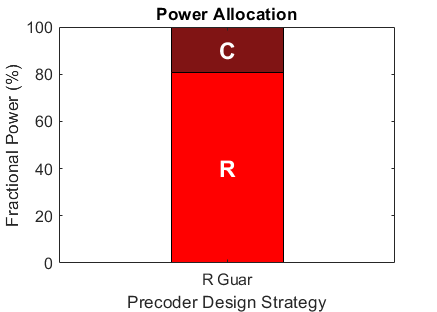}}
\hfil
\subfloat{\includegraphics[width=4.5cm]{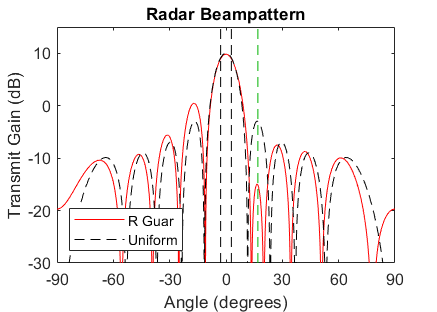}}
\subfloat{\includegraphics[width=4.5cm]{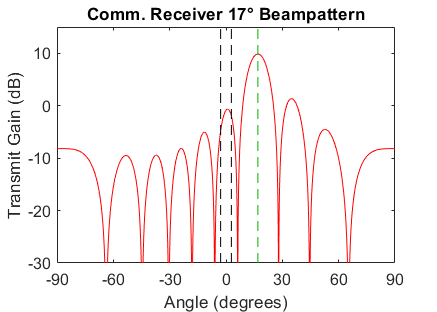}}
\caption{Radar guarantee with greater excess power.}
\label{fig:rguar2}
\end{figure}

If we instead specify a worst-case input SNR of $-33$ dB, the system has much more flexibility, yielding the results in Fig.~\ref{fig:rguar2}.
Now, the system allots more power to the communication subsystem, and the radar beampattern has greatly reduced transmit gain towards the secondary node. It now achieves a communication SINR of 17.4 dB.

\subsection{Communication Guarantee}
As an alternative to the radar guarantee approach, we can guarantee a desired communication SINR for all secondary nodes, while maximizing the radar SINR over discrete search sector $\Theta$ for worst-case target input SNR $\xi_{ir}$. This method, termed \textit{communication guarantee}, is in part motivated by communication systems with error correction coding in which a threshold SNR for error-free performance exists. The starting optimization is identical to \eqref{eq:orig_rguar}, except that $\Gamma_c$ is a constant and $\Gamma_r$ is a variable to be maximized. As with radar guarantee, we use the following iterative approach:
\begin{enumerate}
\item Determine a range of potential values for $\Gamma_r$, the smallest $\Gamma_{r,min}$ being a chosen minimum value, and the largest $\Gamma_{r,max}$ determined from available power. As explained in the radar guarantee example above, we use $\Gamma_{r,max}=N_rM^3\xi_{ir}$.
\item Perform a bisection search, fixing values of $\Gamma_r$ and for each attempting to solve the following problem:
\begin{subequations}
\label{eq:cguar}
\begin{alignat}{2}
\label{eq:cguar1}
&\max_{\bm{R}, \{\bm{R_k}\}_{k=1}^{K_c}, t} &&t \\
\label{eq:cguar2}
&\text{subject to } \quad && \bm{R} \in \mathcal{S}_M^{+}, \\
\label{eq:cguar3}
& && \bm{R_k} \in \mathcal{S}_M^{+}, \quad k=1, \ldots, K_c, \\
\label{eq:cguar4}
& && \bm{R} - \sum_{k=1}^{K_c} \bm{R_k} \in \mathcal{S}_M^{+}, \\
\label{eq:cguar5}
& && [\bm{R}]_{m,m} = 1, \quad m=1, \ldots, M, \\
\label{eq:cguar6}
& && \xi_{oc}'(\bm{R},\bm{R_k},k) \ge \Gamma_c, \quad k=1, \ldots, K_c, \\
\label{eq:cguar7}
& && \xi_{or}'\bigg{(}\bm{R},\sum_{k=1}^{K_c} \bm{R_k},\theta,\xi_{ir}\bigg{)} \ge \Gamma_r, \quad \forall \theta \in \Theta, \\
\nonumber
& && M\xi_{ir}\bm{a}^H(\theta)\bigg{(}N_r\bm{R}-(N_r+\Gamma_r)\sum_{k=1}^{K_c}\bm{R_k}\bigg{)} \\
\label{eq:cguar8}
& && \quad \cdot \bm{a}(\theta) - \Gamma_r \ge t, \quad \forall \theta \in \Theta.
\end{alignat}
\end{subequations}
Constraint \eqref{eq:cguar8}, which derives from a rearrangement of \eqref{eq:R_xi_or}, serves to push us closer to the globally optimal value for $\Gamma_r$.
\item Start the search with $\Gamma_r = \Gamma_{r,min}$. If \eqref{eq:cguar} is feasible, continue, and in successive steps, choose $\Gamma_r$ as the midpoint of the remaining search range. Otherwise, terminate.
\item At each search step, if \eqref{eq:cguar} is feasible, set the used value of $\Gamma_r$ as the new lower limit, and if it is infeasible, as the new upper limit. Stop once the search has narrowed to a range of at most a user-specified $\epsilon_r$.
\item Denoting the optimal solution to \eqref{eq:cguar} with $\Gamma_r$ set to be the lower limit on the final search range by $\widetilde{\bm{R}}$, $\{\widetilde{\bm{R_k}}\}_{k=1}^{K_c}$, $\widetilde{t}$, form the columns of $\bm{W_c}$ as
\begin{equation}
\label{eq:cguar_wk}
\bm{w_k} = \Big{(}\bm{h_k}^H\widetilde{\bm{R_k}}\bm{h_k}\Big{)}^{-1/2}\widetilde{\bm{R_k}}\bm{h_k},\quad k=1,\ldots,K_c.
\end{equation}
\item Choose $\bm{W_r}$ such that
\begin{equation}
\label{eq:cguar_Wr}
\bm{W_r}\bm{W_r}^H = \widetilde{\bm{R}}-\bm{W_c}\bm{W_c}^H.
\end{equation}
We perform this via eigen-decomposition.
\end{enumerate}
As with radar guarantee, the precoder recovered by communication guarantee achieves minimum radar SINR at most $\epsilon_r$ below the optimal value.

\begin{figure}[!t]
\centering
\subfloat{\includegraphics[width=4.5cm]{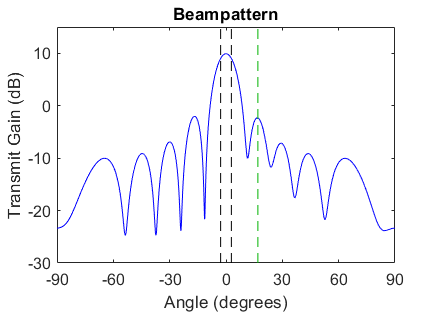}}
\subfloat{\includegraphics[width=4.5cm]{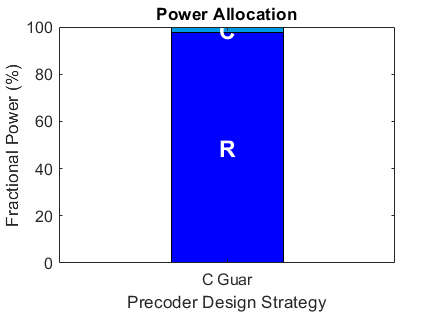}}
\hfil
\subfloat{\includegraphics[width=4.5cm]{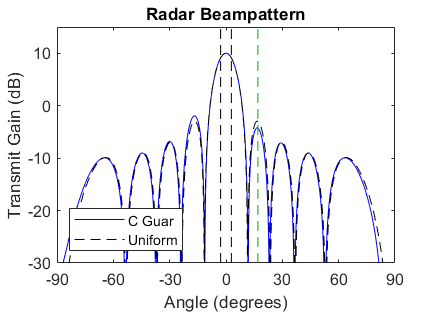}}
\subfloat{\includegraphics[width=4.5cm]{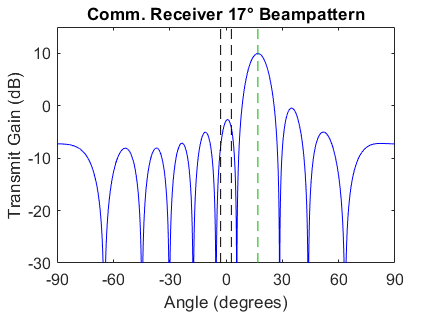}}
\caption{Communication guarantee with easily met communication constraint.}
\label{fig:cguar1}
\end{figure}

To see this method's functionality, we analyze the same example as that used to explore the radar guarantee method above. We require 5 dB of communication SINR, and we use $-33$ dB for the worst-case target input SNR. Here and in all future examples, we do our bisection search in log-space with $\epsilon_r=0.1$ dB. The results are shown in Fig.~\ref{fig:cguar1}.
We see that nearly all power is allocated to the radar subsystem, since very little power is required by the communications subsystem to achieve 5 dB of communication SINR at the single secondary node. In the radar beampattern plot, we again plot the beampattern of a uniform beamformer for comparison. We see that the two curves are very similar, with the communication guarantee radar beampattern having slightly lower transmit gain towards the secondary node.

\begin{figure}[!t]
\centering
\subfloat{\includegraphics[width=4.5cm]{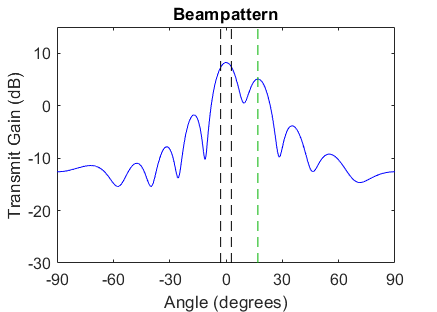}}
\subfloat{\includegraphics[width=4.5cm]{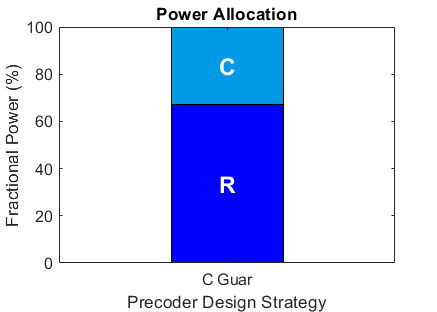}}
\hfil
\subfloat{\includegraphics[width=4.5cm]{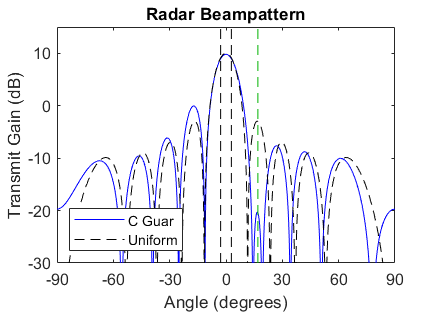}}
\subfloat{\includegraphics[width=4.5cm]{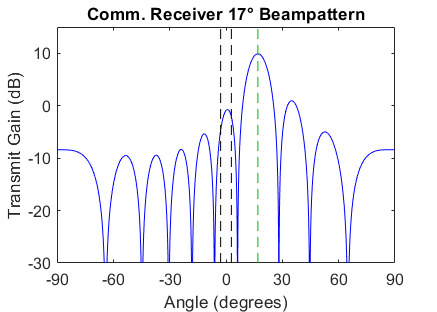}}
\caption{Communication guarantee with more straining communication constraint.}
\label{fig:cguar2}
\end{figure}

If we instead specify a 20 dB communication SINR guarantee, we get the results shown in Fig.~\ref{fig:cguar2}.
Now, much more power is allotted to the communication subsystem (33\% rather than 2.2\%), and the radar beampattern has greatly reduced transmit gain towards the secondary node, as the system now needs to meet a much higher communication SINR threshold.

\subsection{Number of Radar Waveforms}
We now compare the radar and communication guarantee methods by how many radar waveforms are required by each. Many traditional radar systems utilize a single radar waveform, transmitted via a transmit beamformer from an array antenna. Other MIMO radar systems instead utilize multiple radar waveforms, transmitted using multiple transmit beamformers from an array antenna. We strive for our methods to be applicable to both types of systems. The radar transmit covariance matrices that result from solving radar and communication guarantee, shown in \eqref{eq:rguar_Wr} and \eqref{eq:cguar_Wr}, are in general rank $M$, suggesting the use of $M$ radar waveforms. However, we have found empirically that depending on the choice of search sector $\Theta$, fewer radar waveforms may be required.

\begin{figure}[!t]
\centering
\subfloat{\includegraphics[width=4.5cm]{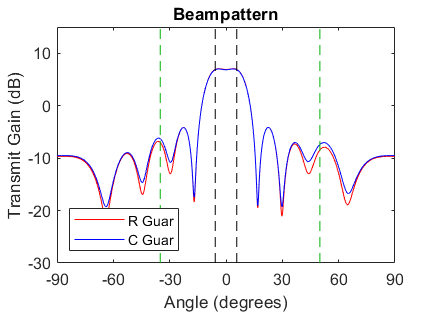}}
\subfloat{\includegraphics[width=4.5cm]{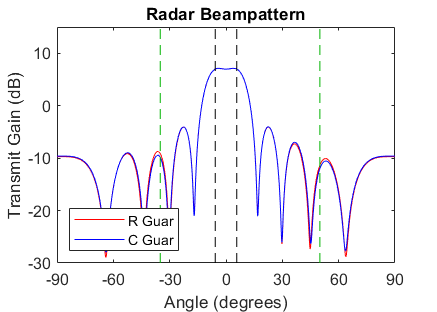}}
\hfil
\subfloat{\includegraphics[width=4.5cm]{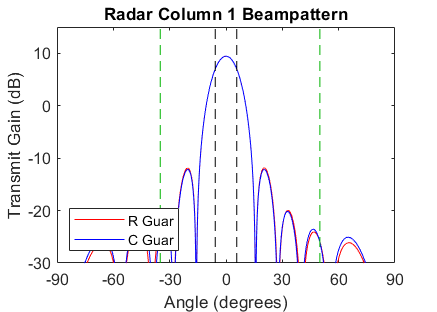}}
\subfloat{\includegraphics[width=4.5cm]{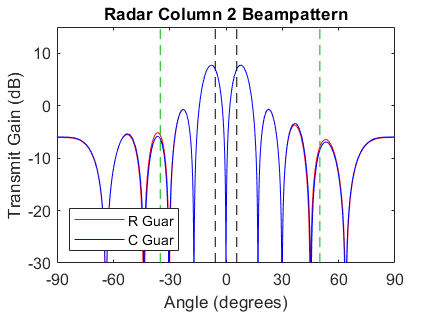}}
\hfil
\subfloat{\includegraphics[width=4.5cm]{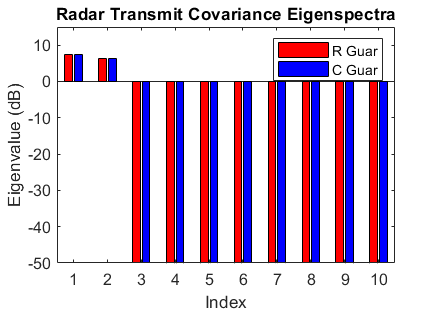}}
\caption{Beamwidth search sector.}
\label{fig:num_radar_waveforms_beamwidth}
\end{figure}

To see this, we first simulate a scenario where $\Theta$ (which we refer to as the \textit{search sector}) is a beamwidth centered at broadside, discretized in increments of one-tenth of a beamwidth in sine space. The radar guarantee approach is constrained to achieve at least 15 dB radar SINR for a target with worst-case input SNR $-31.9$ dB, using a length-100 radar pulse. Here and in future examples, we use the same worst-case target input SNR for radar and communication guarantee. We place two secondary nodes, one at $-35$\textdegree, and one at 50\textdegree, each with input SNR $-5$ dB, and the communication guarantee approach is constrained to achieve at least 5 dB communication SINR using a length-10 communication symbol pulse shape. The results are shown in Fig.~\ref{fig:num_radar_waveforms_beamwidth}.
We see from the radar transmit covariance matrix eigenspectra plot that both approaches essentially put all radar power into two waveforms, as these eigenvalues correspond to relative system power allocations to the different radar waveforms.

The waveform receiving more power is associated with the column of $\bm{W_r}$ which defines the radar column 1 beampattern, which has its mainlobe centered with the search sector. The waveform receiving less power is associated with the column of $\bm{W_r}$ which defines the radar column 2 beampattern, which  has a null in the search sector but two high lobes at its edges. The composite results in radar beampatterns, and in turn total beampatterns, that are fairly flat across the search sector.

\begin{figure}[!t]
\centering
\subfloat{\includegraphics[width=4.5cm]{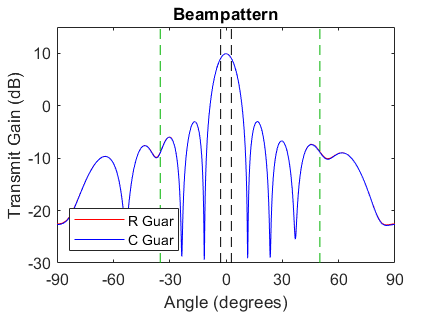}}
\subfloat{\includegraphics[width=4.5cm]{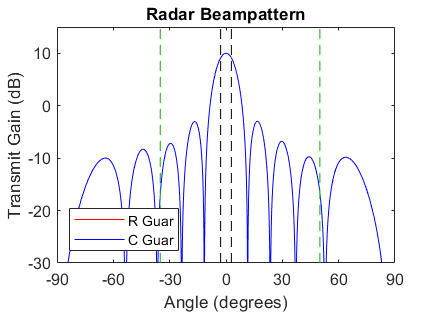}}
\hfil
\subfloat{\includegraphics[width=4.5cm]{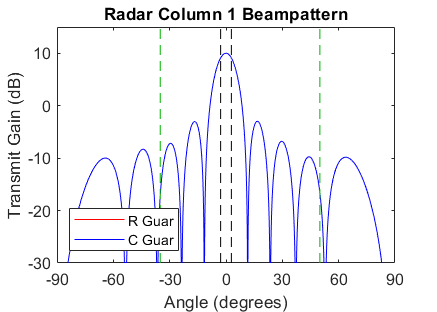}}
\subfloat{\includegraphics[width=4.5cm]{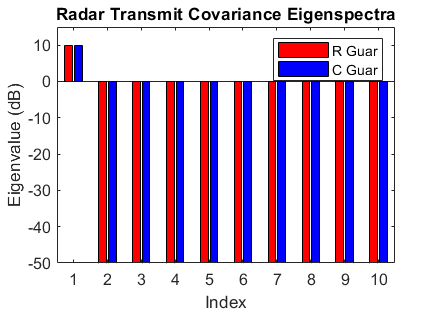}}
\caption{Half-beamwidth search sector.}
\label{fig:num_radar_waveforms_half_beamwidth}
\end{figure}

Now, we change our search sector $\Theta$ to have width half of a beamwidth. We also change our worst-case target input SNR to $-34$ dB since we are searching a smaller space, allowing us to achieve comparable radar SINRs for targets with lower input SNRs under the same system power constraints. The results are shown in Fig.~\ref{fig:num_radar_waveforms_half_beamwidth}.
We see that essentially all radar power is put into a single waveform, meaning one radar waveform could be used instead of 10 with minimal degradation. In addition, the full beamwidth centered at broadside is covered fairly well, meaning a single radar waveform could be used to cover the whole beamwidth without major losses.

\subsection{Power Tradeoffs}
\label{ss:power}
We next compare our methods in terms of power allocations between subsystems and communication SINRs achieved. We simulate a scenario similar to that above, searching half of a beamwidth. The radar guarantee approach is constrained to achieve at least 15 dB radar SINR for a target with worst-case input SNR $-34$ dB, using a length-100 radar pulse. We place a single secondary node at 30\textdegree\, with input SNR $-12$ dB, and the communication guarantee approach is constrained to achieve at least 5 dB communication SINR using a length-10 communication symbol pulse shape.

\begin{figure}[!t]
\centering
\subfloat{\includegraphics[width=4.5cm]{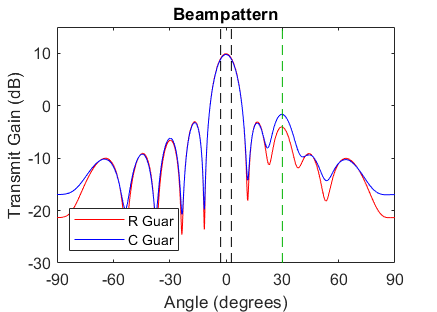}}
\subfloat{\includegraphics[width=4.5cm]{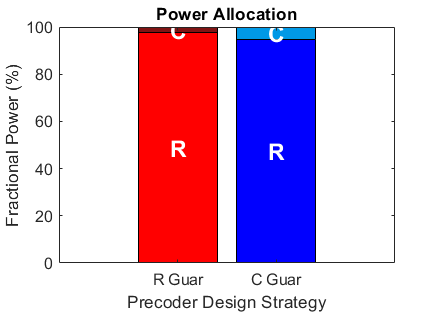}}
\hfil
\subfloat{\includegraphics[width=4.5cm]{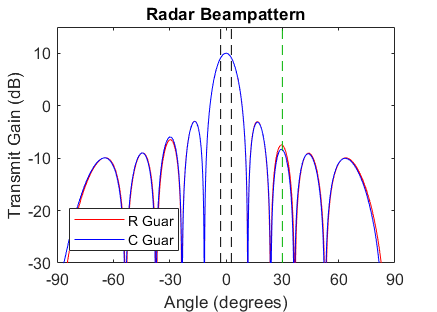}}
\subfloat{\includegraphics[width=4.5cm]{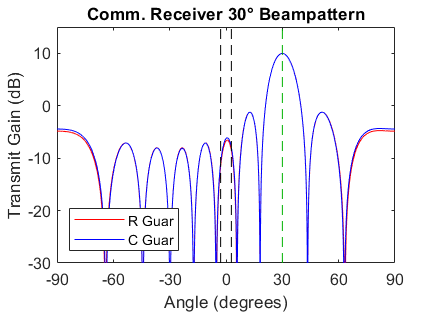}}
\hfil
\subfloat{\includegraphics[width=4.5cm]{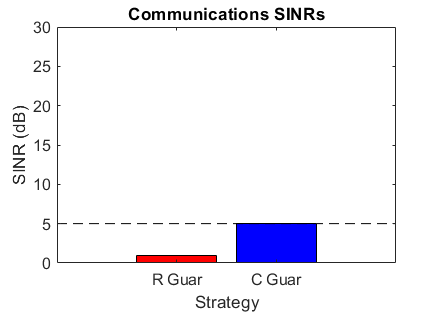}}
\caption{Power-constrained comparison.}
\label{fig:rc_comp1_hard}
\end{figure}

Since we search half of a beamwidth, both here and in coming examples, we use a single radar waveform, as we would like to see that we can indeed use just one for this purpose. To accomplish this for these two methods, we use eigen-decomposition to form a single-column radar precoder from a rank-one approximation of $\widetilde{\bm{R}}-\bm{W_c}\bm{W_c}^H$, as found in \eqref{eq:rguar_Wr} and \eqref{eq:cguar_Wr}. We then renormalize all rows of $\bm{W}$ to meet our power constraint with equality. The results are shown in Fig.~\ref{fig:rc_comp1_hard}.

We see that the radar and communication beampatterns are similar between the two approaches. This trend of similar column beampatterns persists through our remaining examples. The total beampatterns are also similar, though the radar guarantee beampattern has slightly higher transmit gain in the search sector, while the communication guarantee beampattern has higher transmit gain towards the secondary node. We also see that while communication guarantee achieves 5 dB communication SINR, radar guarantee achieves only 1 dB SINR. This is because this is a power-constrained setup, in which there are insufficient power resources available to meet both the radar and communication performance thresholds.

\begin{figure}[!t]
\centering
\subfloat{\includegraphics[width=4.5cm]{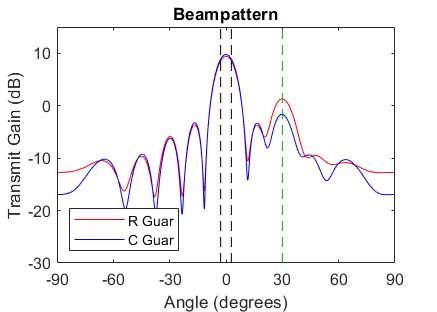}}
\subfloat{\includegraphics[width=4.5cm]{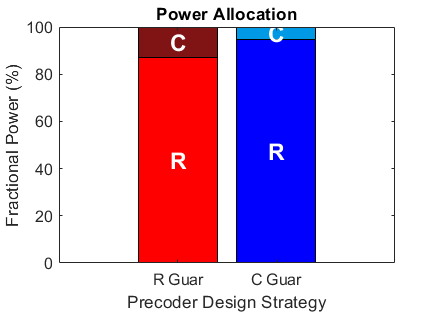}}
\hfil
\subfloat{\includegraphics[width=4.5cm]{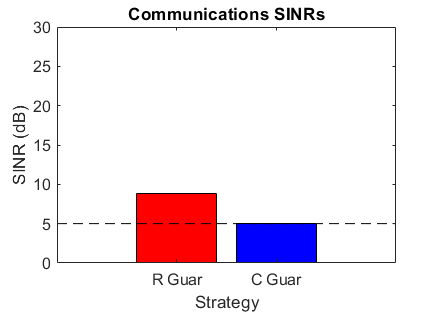}}
\caption{Excess power comparison.}
\label{fig:rc_comp1_easy}
\end{figure}

However, if we change the worst-case input SNR to be $-33.5$ dB, the setup has excess power and yields the results shown in Fig.~\ref{fig:rc_comp1_easy}.
Now, radar guarantee has higher transmit gain towards the secondary node and achieves a communication SINR of 8.8 dB, above the 5 dB threshold, while communication guarantee has higher transmit gain within the search sector than radar guarantee. Radar guarantee now allots less power for the radar subsystem than does communication guarantee. We see from this that depending on available power resources, either method may prioritize either subsystem.

\begin{figure}[!t]
\centering
\subfloat{\includegraphics[width=4.5cm]{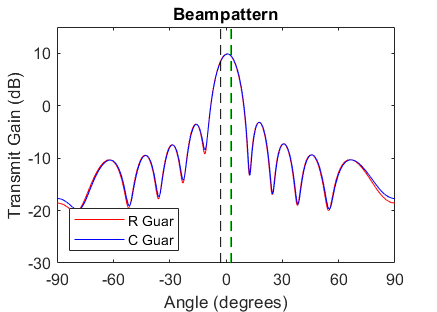}}
\subfloat{\includegraphics[width=4.5cm]{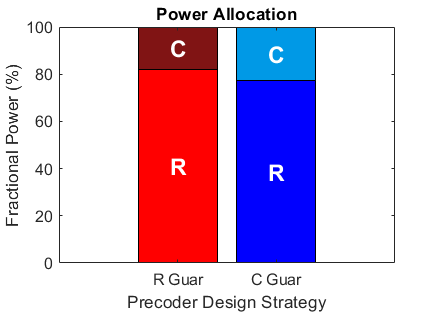}}
\hfil
\subfloat{\includegraphics[width=4.5cm]{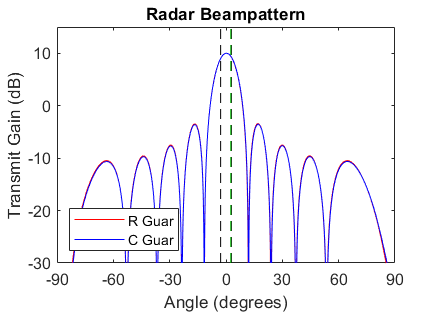}}
\subfloat{\includegraphics[width=4.5cm]{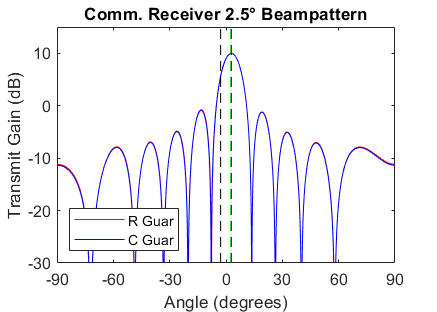}}
\hfil
\subfloat{\includegraphics[width=4.5cm]{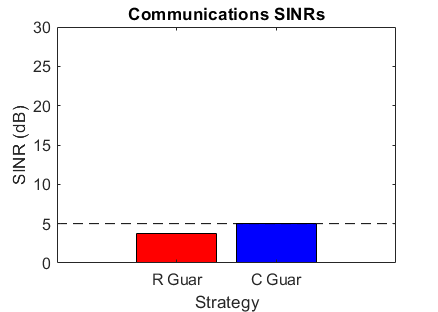}}
\caption{Secondary node in search sector comparison.}
\label{fig:rc_comp_inside}
\end{figure}

As a further example, we see what happens when a secondary node is within the search sector. Specifically, we place a single secondary node at 2.5\textdegree\ with input SNR $-5$ dB. The communication guarantee is 5 dB, while the radar guarantee is 15 dB for worst-case input SNR $-33$ dB. This yields the results shown in Fig.~\ref{fig:rc_comp_inside}.

Interestingly, the radar beampattern of each method is centered just to the right of broadside (though this is difficult to see visually, the peak transmit gain of each is around 0.2\textdegree). This is because both methods are trying to offset the communications power transmitted in the right half of the search sector. The communication beampattern of each method is also centered a bit to the right of the secondary node, as both methods are trying to reduce communications transmit gain within the search sector.

\begin{figure}[!t]
\centering
\subfloat{\includegraphics[width=4.5cm]{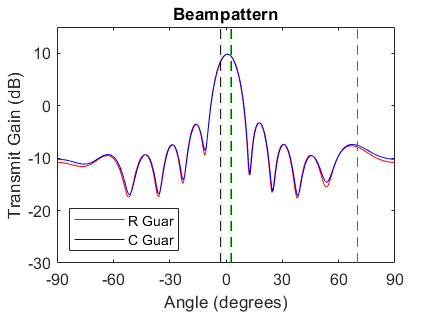}}
\subfloat{\includegraphics[width=4.5cm]{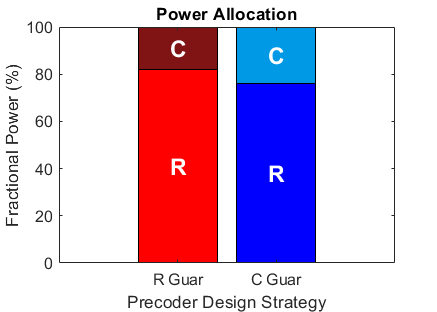}}
\hfil
\hfil
\subfloat{\includegraphics[width=4.5cm]{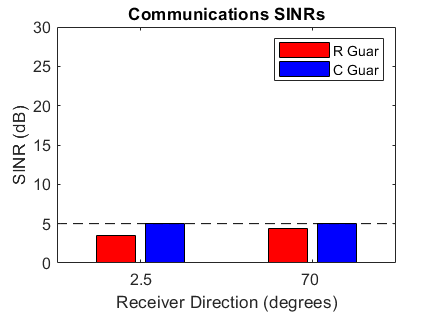}}
\caption{One of two secondary nodes in search sector comparison.}
\label{fig:rc_comp_foreshadow}
\end{figure}

As a final example, we add in an additional secondary node at 70\textdegree, also with $-5$ dB input SNR, and the results are shown in Fig.~\ref{fig:rc_comp_foreshadow}.
We see similar results as in the previous example. Notably, despite allotting a fair amount of power to the communication subsystem, radar guarantee does not achieve the 5 dB communication SINR threshold at either secondary node. This is because the method is designed to maximize the minimum communication SINR across all secondary nodes, so because it cannot achieve a 5 dB SINR at the secondary node at 2.5\textdegree, it does not try to achieve an SINR that high at the other secondary node, either. This observation motivates a new method explored in Section~\ref{section:priority}.

We see from the above examples that radar and communication guarantee provide for specifying a desired performance level for either subsystem, while allocating all remaining resources to the other. Due to the flexibility of channel and input SNR specifications, as well as desired SINR performance levels, these methods could be applicable in a variety of settings, such as surveillance settings in which the radar is searching at long ranges and the communication system may be operating over short or long distances, or automotive settings in which the radar is operating over short ranges and the communication system is operating over longer ranges.

\section{Radar Priority Approach}
\label{section:priority}
\noindent In this section, we introduce a new method that addresses the undesirable behavior observed in the radar guarantee approach, namely that a single difficult-to-service secondary node can prevent the closure of any communication link. This motivates a new approach which tries to maximize the number of communication links closed, rather than maximizing the minimum communication SINR.

As such, we aim to solve the following problem:
\begin{subequations}
\label{eq:orig_rpriority}
\begin{alignat}{2}
\label{eq:orig_rpriority1}
&\max_{\bm{W_c}, \bm{W_r}, \mathcal{K}} && |\mathcal{K}| \\
\label{eq:orig_rpriority2}
&\text{subject to } \quad && \bm{R}=\bm{W_c}\bm{W_c}^H+\bm{W_r}\bm{W_r}^H, \\
\label{eq:orig_rpriority3}
& && [\bm{R}]_{m,m} = 1, \quad m=1, \ldots, M, \\
\label{eq:orig_rpriority4}
& && \mathcal{K} \subseteq \{1, \ldots, K_c\}, \\
\label{eq:orig_rpriority6}
& && \xi_{oc}(\bm{W_c},\bm{W_r},k)\ge\Gamma_c, \quad \forall k \in \mathcal{K}, \\
\label{eq:orig_rpriority5}
& && \xi_{or}(\bm{W_c},\bm{W_r},\theta,\xi_{ir}) \ge \Gamma_r, \quad \forall \theta \in \Theta.
\end{alignat}
\end{subequations}
Here, $|\mathcal{K}|$ denotes the cardinality of set $\mathcal{K}$. In contrast with radar guarantee, both $\Gamma_r$ and $\Gamma_c$ are user-specificed inputs.

There may be a range of optimal solutions to this problem, since there may be excess power after satisfying the radar constraint and closing however many communication links can be closed. Therefore, after determining which secondary nodes to service, we run the communication guarantee approach specifying only these secondary nodes. This allots all excess power, after closing as many communication links as possible under the radar guarantee, to the radar subsystem. We term this the \textit{radar priority} approach. The remaining task is solving \eqref{eq:orig_rpriority}. There are multiple approaches we could take.

\subsection{Combinatorial Approach}
One approach is to try every combination of secondary nodes. Specifically, we could use the following process:
\begin{enumerate}
\item Sequentially specify each subset $\mathcal{K}$ of secondary nodes.
\item For each $\mathcal{K}$, check if the following problem is feasible:
\begin{subequations}
\label{eq:rpriority_comb}
\begin{alignat}{2}
\label{eq:rpriority_comb1}
&\text{Find } && \bm{R}, \{\bm{R_k}\}_{k\in\mathcal{K}} \\
\label{eq:rpriority_comb2}
&\text{subject to } \quad && \bm{R} \in \mathcal{S}_M^{+}, \\
\label{eq:rpriority_comb3}
& && \bm{R_k} \in \mathcal{S}_M^{+}, \quad k\in\mathcal{K}, \\
\label{eq:rpriority_comb4}
& && \bm{R} - \sum_{k\in\mathcal{K}} \bm{R_k} \in \mathcal{S}_M^{+}, \\
\label{eq:rpriority_comb5}
& && [\bm{R}]_{m,m} = 1, \quad m=1, \ldots, M, \\
\label{eq:rpriority_comb7}
& && \xi_{oc}'(\bm{R},\bm{R_k},k) \ge \Gamma_c, \quad \forall k \in \mathcal{K}, \\
\label{eq:rpriority_comb6}
& && \xi_{or}'\bigg{(}\bm{R},\sum_{k\in\mathcal{K}} \bm{R_k},\theta,\xi_{ir}\bigg{)} \ge \Gamma_r, \quad \forall \theta \in \Theta.
\end{alignat}
\end{subequations}
\item Choose the subset $\mathcal{K}$ of largest cardinality for which \eqref{eq:rpriority_comb} is feasible. We break ties by choosing the subset which, after using the communication guarantee approach, requires the least amount of power for the communication subsystem. We note that if a non-square radar precoder is going to be formed as explained at the beginning of subsection~\ref{ss:power}, then this is done as part of solving communication guarantee for the tie-breaking done.
\end{enumerate}
The approach above is guaranteed to find the optimal $\mathcal{K}$ in \eqref{eq:orig_rpriority}, since every possible choice for $\mathcal{K}$ is tried. By Theorem~\ref{thm:cov_iff}, \eqref{eq:rpriority_comb} is feasible if and only if precoders exist satisfying the per-antenna power constraint, the radar guarantee constraint, and achieving $\Gamma_c$ communications SINR for each secondary node in $\mathcal{K}$. However, this process takes exponential time in the number of secondary nodes.

\subsection{Greedy Algorithm Approach}
While the combinatorial approach finds the globally optimal solution, its runtime is asymptotically slow. We therefore propose an algorithm with a runtime polynomial in the number of secondary nodes, though it is not guaranteed to find the globally optimal solution. The general approach is to first rank secondary nodes in terms of `difficulty to service', and then to utilize this order in choosing easier secondary nodes first in trying to service as many as possible.

To accomplish this, for each $k=1,\ldots,K_c$, we attempt to solve \eqref{eq:rpriority_comb} with $\mathcal{K}=\{k\}$. If this problem is feasible, it means that we can close the $k$-th communication link subject to our radar guarantee, and we add this secondary node to set $\mathcal{K}_{feas}$. For each secondary node in $\mathcal{K}_{feas}$, we run the communication guarantee approach, specifying only that single secondary node, and we see how much power is needed by the system for communications. Once again, choosing a smaller radar precoder would be done as part of communication guarantee at this step. After doing this for all secondary nodes, we rank them from least to most power required for communications, forming the ordered list of feasible secondary nodes $\mathcal{K}_{feas}^{ord}$.

Next, we initialize an empty set $\mathcal{K}_{greedy}$. We then iterate through $\mathcal{K}_{feas}^{ord}$, adding one secondary node at a time to $\mathcal{K}_{greedy}$ and seeing if \eqref{eq:rpriority_comb} is feasible using $\mathcal{K} = \mathcal{K}_{greedy}$. If it is, then we keep the most recently added secondary node. If it is not, we discard it. We continue until we have iterated through the entirety of $\mathcal{K}_{feas}^{ord}$. The final version of $\mathcal{K}_{greedy}$ serves as our set of secondary nodes to service.

\subsection{Numerical Results}
\begin{figure}[!t]
\centering
\subfloat{\includegraphics[width=4.5cm]{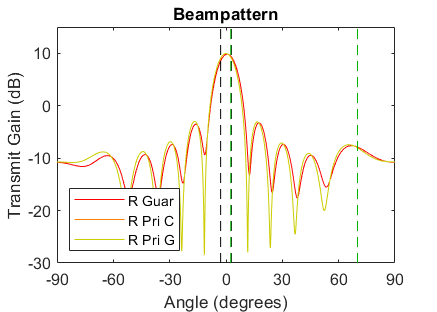}}
\subfloat{\includegraphics[width=4.5cm]{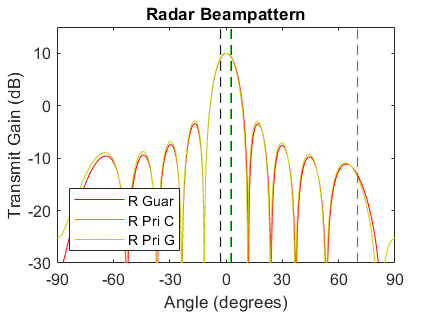}}
\hfil
\subfloat{\includegraphics[width=4.5cm]{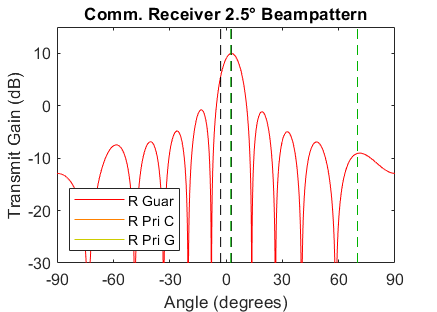}}
\subfloat{\includegraphics[width=4.5cm]{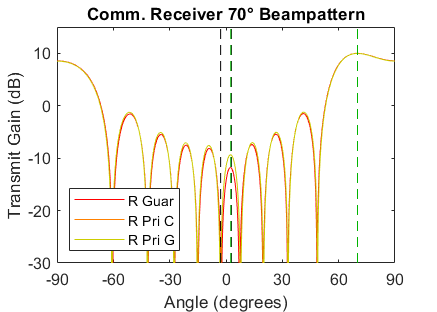}}
\hfil
\subfloat{\includegraphics[width=4.5cm]{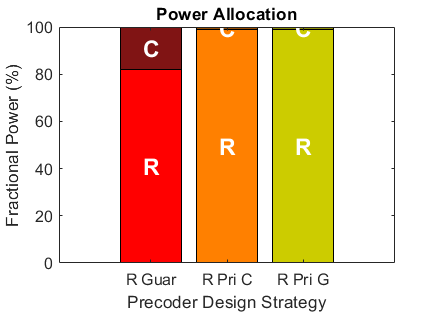}}
\subfloat{\includegraphics[width=4.5cm]{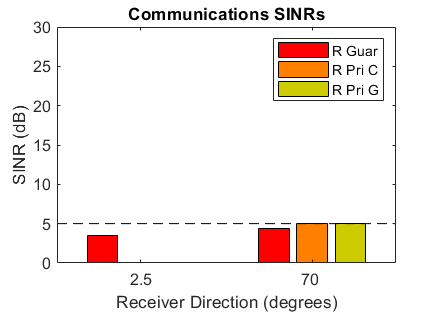}}
\caption{Radar priority versus radar guarantee with secondary node in search sector.}
\label{fig:rpriority_rcguar_comp_ex}
\end{figure}

For our first example, we consider the same scenario as that considered at the end of Section~\ref{section:guar}. The communication SINR threshold for a closed link is 5 dB. Here and in other examples in this section, we again use a single radar waveform. The results are shown in Fig.~\ref{fig:rpriority_rcguar_comp_ex}, with the combinatorial and greedy versions of radar priority denoted as `R Pri C' and `R Pri G', respectively. We note that here and in subsequent communication SINR plots, no value reported for radar priority indicates that no transmissions were made for the corresponding secondary node, yielding 0 SINR.

As we saw above, radar guarantee does not close either communication link, achieving communication SINRs of roughly 3.5 and 4.3 dB, while using 18.1\% of system power for communications. Meanwhile, both radar priority versions close the link to the secondary node at 70\textdegree\ using only 1.1\% of system power for communications. The two versions of radar priority yield the same results, as they both reduce to communication guarantee with only the secondary node at 70\textdegree\ specified. We see that all beampatterns look fairly similar between methods, though only radar guarantee has a nonzero beampattern for the secondary node at 2.5\textdegree. This similarity in beampatterns continues through our remaining examples.

\begin{figure}[!t]
\centering
\subfloat{\includegraphics[width=4.5cm]{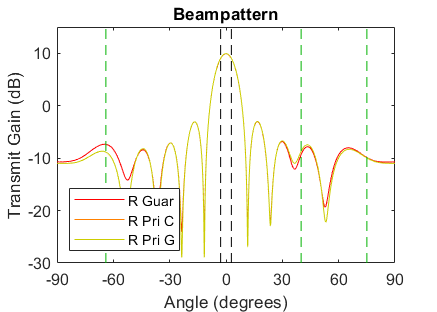}}
\subfloat{\includegraphics[width=4.5cm]{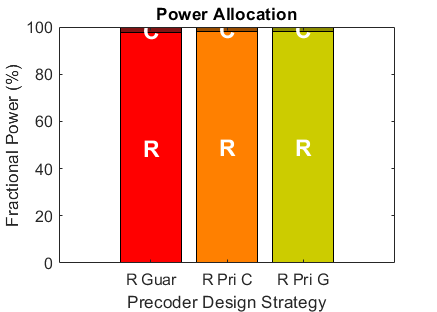}}
\hfil
\subfloat{\includegraphics[width=4.5cm]{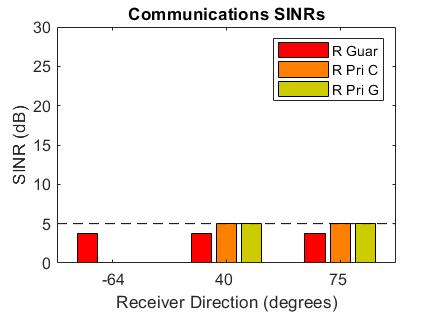}}
\caption{Power limited comparison.}
\label{fig:radar_priority_power_limited}
\end{figure}

A different situation that brings out the potentially undesirable behavior in radar guarantee is when no individual secondary node is particularly difficult to service, but power resources do not allow for all to be serviced. For instance, we consider a situation with the same search sector, with three secondary nodes at $-64$\textdegree, $40$\textdegree, and $75$\textdegree, all with $-4$ dB input SNR. We make the worst-case target input SNR $-34$ dB. The results are shown in Fig.~\ref{fig:radar_priority_power_limited}.
In this situation, radar guarantee does not close any of the three communication links, achieving communication SINRs of roughly $3.7$ dB for all three secondary nodes. The communication subsystem in this case uses $2.2$\% of the available power. However, both versions of the radar priority approach, which again yield the same results as one another, close links with the secondary nodes at $40$\textdegree\ and $75$\textdegree, using about $1.8$\% of system power.

From these examples, we see that radar priority is able to close links in scenarios where radar guarantee is able to close none, while using less power for communications. The combinatorial version is guaranteed to close as many links as possible, though its asymptotic runtime is exponential in the number of secondary nodes. The greedy version is often able to find the largest possible set of links to close, though this is not guaranteed. We note that in a setting where all communication links can be closed under the radar constraint, both versions of radar priority are equivalent to communication guarantee.

\section{Comparison to Prior Methods}
\label{section:prior_methods}
\noindent In this section, we describe prior methods found in the literature and compare them to our approaches.

\subsection{Previous Methods}
The work \cite{precoder_paper} considers a similar scenario to ours but defines a metric $L(\bm{R},\alpha)$, dependent on the waveform covariance matrix $\bm{R}=\bm{W}\bm{W}^H$, as a weighted sum of two terms:
\begin{enumerate}
\item MSE between the directional power transmitted and some desired such pattern scaled by parameter $\alpha$;
\item Cross-correlation between radar directions of interest.
\end{enumerate}
For the first term, with sampled angle grid $\{\theta_q\}_{q=1}^{Q}$ and $d(\theta)$ a desired transmitted power level in direction $\theta$, \cite{precoder_paper} defines
\begin{equation*}
L_1(\bm{R},\alpha) = \frac{1}{Q}\sum_{q=1}^{Q}\Big{|}\alpha d(\theta_q) - P_e\bm{a}^H(\theta_q)\bm{R}\bm{a}(\theta_q)\Big{|}^2.
\end{equation*}
For the second term, letting $\{\bar{\theta}_q\}_{q=1}^{Q_{tgt}}$ denote a collection of target directions of interest, it sets
\begin{equation*}
L_2(\bm{R}) = \frac{2}{Q_{tgt}^2-Q_{tgt}}\sum_{q_1=1}^{Q_{tgt}-1}\sum_{q_2=q_1+1}^{Q_{tgt}}\Big{|}P_e\bm{a}^H(\bar{\theta}_{q_2})\bm{R}\bm{a}(\bar{\theta}_{q_1})\Big{|}^2.
\end{equation*}
It then defines $L(\bm{R},\alpha) = L_1(\bm{R},\alpha)+wL_2(\bm{R})$, where $w$ is some chosen weighting. We note that in \cite{precoder_paper}, the $P_e$ factors are not shown in these metrics since we have factored $P_e$ out of the transmit covariance matrices.

The first strategy in \cite{precoder_paper}, termed the \textit{MSE} approach, minimizes $L(\bm{R},\alpha)$ for arbitrary $\alpha$, subject to a per-antenna power constraint and a communication SINR constraint. The other strategy, termed the \textit{zero forcing} (ZF) approach, is similar to the MSE approach, but it additionally eliminates all interference at all secondary nodes\footnote{In \cite{precoder_paper}, the transmit power is incorporated into the precoder. However, since we have included the transmit power elsewhere in our metrics, the formulations are mathematically equivalent. In addition, in \cite{precoder_paper}, the communication SINR constraint is sample- rather than symbol-level. However, this simply changes each secondary node's SINR threshold by a constant scalar and does not affect the structure of the optimization.}.

The radar guarantee and priority approaches differ from the others by guaranteeing radar performance and maximizing communication performance, rather than the reverse. Communication guarantee differs from the MSE and ZF approaches primarily in the radar metric being optimized. The radar SINR metric drives communication guarantee to maximize radar and minimize communication transmit power toward the search sector. The $L(\cdot)$ metric drives the MSE and ZF approaches to try to approximate some desired beampattern, irrespective of whether transmit power is coming from radar or communication waveforms. This distinction causes communication guarantee to rely far more on radar transmissions for radar sensing in scenarios of interest. Since radar waveforms are often chosen to have desirable ambiguity functions for detection purposes, communication guarantee is able to better approximate these desired ambiguity functions than MSE and ZF are. We see that via simulation in the next subsection.

\subsection{Numerical Comparison}
In this subsection, we compare the new and prior methods from the perspective of waveform characteristics in directions of interest, specifically the delay-Doppler characteristics of waveforms captured by their ambiguity functions.

We utilize a single radar waveform. We do this as described above for radar and communication guarantee. For MSE and ZF, we use the returned $\bm{W_r}$ to compute $\bm{W_r}\bm{W_r}^H$. We then form a single-column radar precoder from a rank-1 approximation of $\bm{W_r}\bm{W_r}^H$, and then we renormalize all rows of $\bm{W}$ to meet our power constraint with equality\footnote{With the MSE and ZF approaches, renormalizing only the new radar precoder has been observed to yield precoders which violate the communication SINR guarantee in some scenarios in which the procedure used does not.}.

As in previous examples, we use our half-beamwidth search sector $\Theta$ centered at broadside. For the MSE and ZF approaches, we set the desired beampattern to 1 both within and just outside the half beamwidth centered at broadside, and 0 elsewhere\footnote{Setting to 1 only within a search region of this size and 0 elsewhere has been observed to provide worse coverage in some scenarios of interest.}. Since we look in one general direction, we choose $Q_{tgt}=0$.

The radar pulse used is a normalized complex linear frequency modulated (LFM) chirp of duration 25 $\mu$s, swept from $-500$ kHz to 500 kHz, sampled at 4 MHz (setting $N_r=101$). The communication signaling uses quadrature phase shift keying, with a normalized length-7 root-raised cosine pulse shape with rolloff factor 0.5 and no overlap between symbol pulses. We note that since 7 does not evenly divide 101, the final few samples of the communications baseband waveform within a radar pulse timeframe will be zero.

\begin{figure}[!t]
\centering
\subfloat{\includegraphics[width=4.5cm]{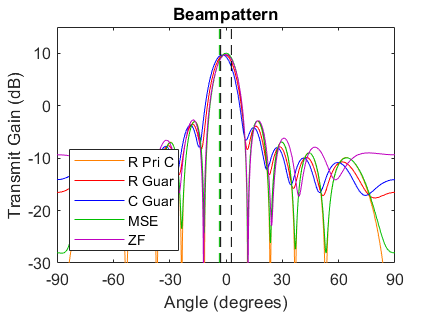}}
\subfloat{\includegraphics[width=4.5cm]{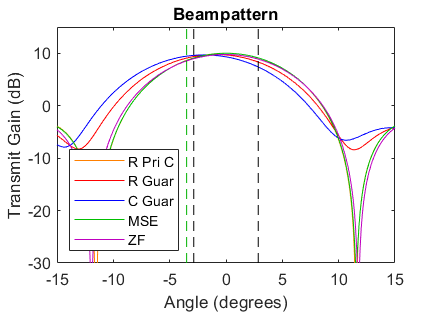}}
\hfil
\subfloat{\includegraphics[width=4.5cm]{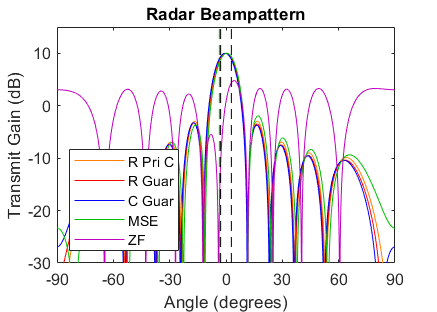}}
\subfloat{\includegraphics[width=4.5cm]{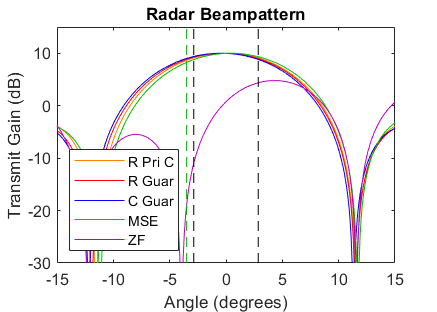}}
\hfil
\subfloat{\includegraphics[width=4.5cm]{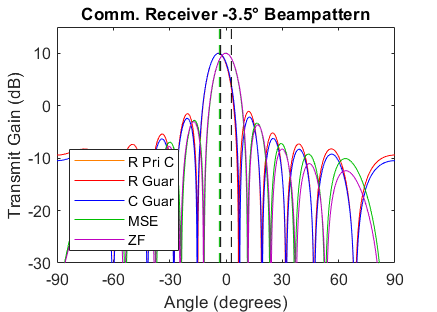}}
\subfloat{\includegraphics[width=4.5cm]{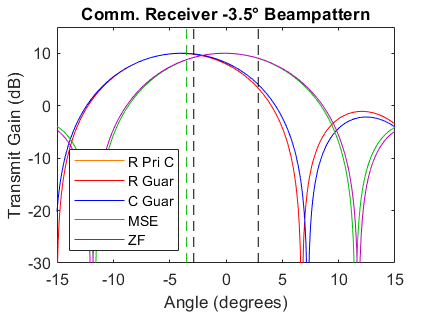}}
\hfil
\subfloat{\includegraphics[width=4.5cm]{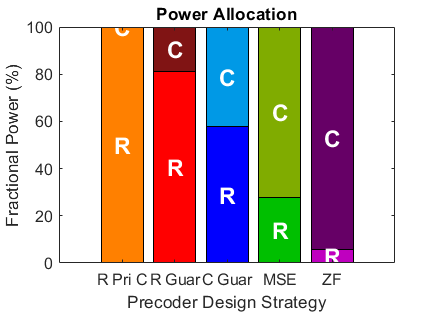}}
\subfloat{\includegraphics[width=4.5cm]{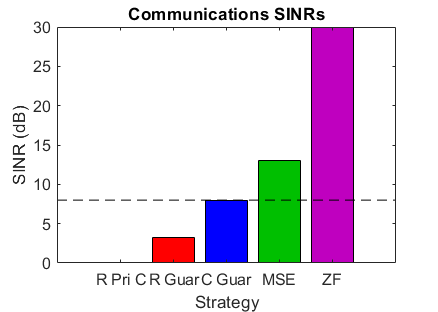}}
\caption{Beampatterns, power, and communication SINR comparison.}
\label{fig:AF1}
\end{figure}

We place one secondary node at $-3.5$\textdegree\ with 3 dB input SNR. We note that this secondary node is very close to the search sector. The communication guarantee is 8 dB SINR per symbol, and the radar guarantee is 15 dB for targets with worst-case input SNRs of $-33$ dB. We place a single target just inside the left boundary of the search sector. The beampattern, power allocation, and communication SINR results are shown in Fig.~\ref{fig:AF1}. We use only the combinatorial version of radar priority since both versions are the same with a single secondary node.

The total beampatterns are similar within the search sector for all five methods. However, we see that while radar and communication guarantee have their communication beampatterns centered right around the secondary node, MSE and ZF center their communication beampatterns with the search sector. Radar priority cannot close the single communication link, so it does not have any communication transmissions. The radar beampatterns are all similar except for that of the ZF approach, which has a null in its radar beampattern around the secondary node\footnote{The location of the null is slighty offset from the secondary node due to the modification done to get a single-column radar precoder.}. The fraction of power allotted to the radar subsystem decreases from radar priority (which allots all power to radar), to radar guarantee, to communication guarantee, to MSE, to ZF. Thus, we observe a stronger reliance on communication transmissions for sensing from the MSE and ZF methods, which achieve higher than required communication SINRs (roughly 13 dB and 29.9 dB, respectively).

\begin{figure}[!t]
\centering
\subfloat{\includegraphics[width=4.5cm]{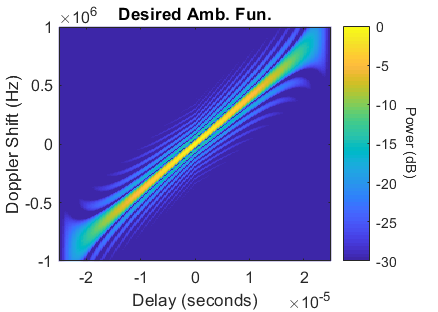}}
\subfloat{\includegraphics[width=4.5cm]{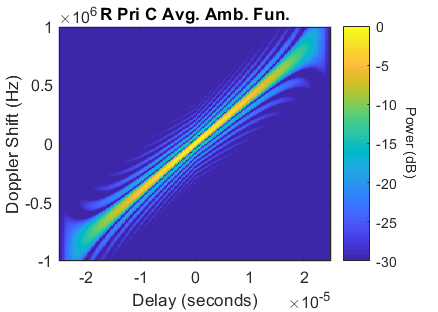}}
\hfil
\subfloat{\includegraphics[width=4.5cm]{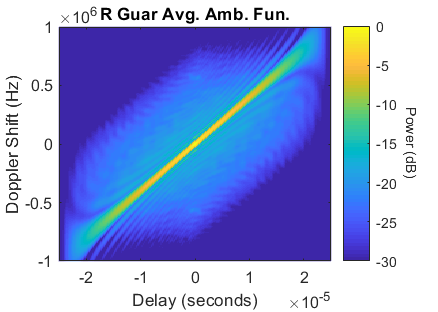}}
\subfloat{\includegraphics[width=4.5cm]{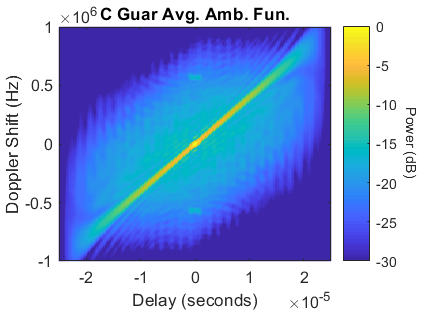}}
\hfil
\subfloat{\includegraphics[width=4.5cm]{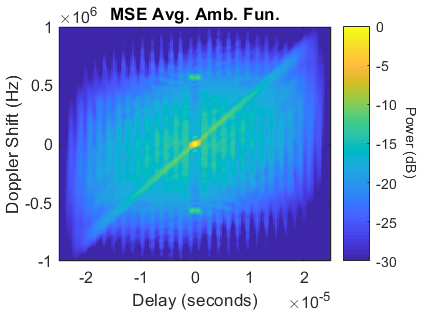}}
\subfloat{\includegraphics[width=4.5cm]{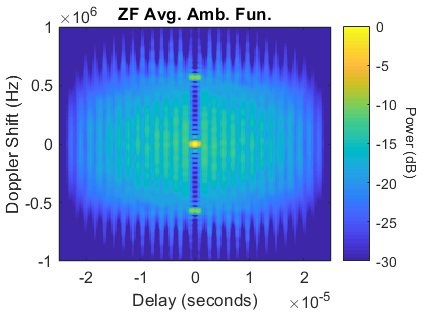}}
\caption{Ambiguity functions comparison.}
\label{fig:AF2}
\end{figure}

For delay $d$ and Doppler shift $f$, we calculate the ambiguity function for length $N$ waveform $x[n]$ as
\begin{equation}
X[d,f] = \bigg{|} \sum_{n=0}^{N-1} \hat{x}[n]\hat{x}^*[n+d]e^{j2\pi n(f/F_s)} \bigg{|},
\end{equation}
where $\hat{x}[n]$ is a unit-norm scaled copy of $x[n]$ and $F_s$ is the sampling rate. To compare our methods, we compute the squares of the ambiguity functions of the waveforms incident on our single target for a single pulse, which by \eqref{eq:incident} are largely determined by the precoder. We do an average over 100 such ambiguity functions, with the differences between them being the communication symbols sent and a random phase factor associated with the channel to the target. The averaged squared magnitudes of the ambiguity functions are shown in Fig.~\ref{fig:AF2}.

The desired ambiguity function is that of the LFM chirp. The ambiguity functions become progressively less similar to that of the chirp as we move from radar guarantee, to communication guarantee, to MSE, to ZF. Since radar priority transmits only the LFM chirp, its ambiguity function is not deteriorated by communications transmissions like the other methods' are. When the communications waveform dominates the target scattered energy, the corresponding ambiguity function more closely resembles that of the communications waveform.

Our original objective was to design a dual-function system utilizing specialized communication and specialized radar waveforms. As such, we constructed our precoder design approaches to rely on communication waveforms for data transmission and radar waveforms for sensing. This comparison of ambiguity functions, which capture the desirable characteristics of the radar waveform chosen, illustrates that the new methods successfully retain these desirable properties to varying degrees, depending on the communications demands.

\section{Conclusion}
\label{section:conclusion}
\noindent We introduced multiple new approaches for transmit precoder design for DFRC systems. These methods are motivated by applications where surveillance radar performance is a priority and there is a desire to multiplex good radar waveforms with good communications waveforms from the same aperture. The techniques developed provide great flexibility in terms of problem setup, desired performance levels, and prioritization of either subsystem. Radar guarantee and priority methods were introduced, the former with a maximin formulation of communication SINR across secondary nodes, the latter maximizing the number of communication links closed, subject to a radar constraint. Communication guarantee was developed as a converse method to radar guarantee. All new approaches were shown to better approximate desired ambiguity functions than prior methods.

In future work, developing new radar and communication metrics to drive the precoder design could be explored. Both incorporating the correlations between the pulse shapes chosen, as well as specifying the metrics to the expected receiver processing to be done, could yield precoders better tailored to specific scenarios. Jointly designing receive processing techniques with the transmit precoders could also yield improved performance. Exploring multi-path channels could further extend the applicability of the methods.

In addition, new approaches could be taken for solving the radar priority method. For instance, developing a mixed integer programming problem with indicator variables for each communication link could lead to different solving methods, such as relaxations to linear programming problems or branch and bound approaches.

Moreover, simultaneously designing a set of precoders for all desired search sectors using something akin to the radar priority approach may be explored. This could involve trying to service each secondary node during some number of search dwells. As the search sector moves around, different secondary nodes may become easier and more difficult to service.

\appendices
\section{Proof of Theorem \ref{thm:cov_iff}}
\label{app:cov_iff}
\noindent We first prove the `if' direction. We assume that there exist precoders $\bm{W_c}$ ($M\times K_c$) and $\bm{W_r}$ ($M\times M$) such that
\begin{equation}
\label{eq:cov_iff_proof_power}
[\bm{W_c}\bm{W_c}^H+\bm{W_r}\bm{W_r}^H]_{m,m} = 1, \quad m=1, \ldots, M,
\end{equation}
\begin{equation}
\label{eq:cov_iff_proof_comm_SINR}
\xi_{oc}(\bm{W_c},\bm{W_r},k) \ge \Gamma_c, \quad k=1, \ldots, K_c,
\end{equation}
and
\begin{equation}
\label{eq:cov_iff_proof_radar_SINR}
\xi_{or}(\bm{W_c},\bm{W_r},\theta,\xi_{ir}) \ge \Gamma_r, \quad \forall \theta \in \Theta.
\end{equation}

With $\bm{W} = [\bm{W_c},\bm{W_r}]$, we choose
\begin{equation*}
\bm{R} = \bm{W}\bm{W}^H, \quad \{\bm{R_k}\}_{k=1}^{K_c} = \bm{w_k}\bm{w_k}^H,
\end{equation*}
where $\bm{w_k}$ denotes the $k$-th column of $\bm{W_c}$. We now verify that these matrices comprise a feasible solution to \eqref{eq:guar}.

Clearly, $\bm{R}$, each $\bm{R_k}$, and $\bm{R}-\sum_{k=1}^{K_c}\bm{R_k} = \bm{W_r}\bm{W_r}^H$ are positive-semidefinite, so \eqref{eq:guar2}, \eqref{eq:guar3}, and \eqref{eq:guar4} are satisfied. By construction of $\bm{R}$, \eqref{eq:cov_iff_proof_power} implies \eqref{eq:guar5}. Next, by definition of $\xi_{oc}'$ and $\xi_{or}'$, we have $\xi_{oc}'(\bm{R},\bm{R_k},k) = \xi_{oc}(\bm{W_c},\bm{W_r},k)$ and $\xi_{or}'\big{(}\bm{R},\sum_{k=1}^{K_c}\bm{R_k},\theta,\xi_{ir}\big{)} = \xi_{or}(\bm{W_c},\bm{W_r},\theta,\xi_{ir}).$ Thus, \eqref{eq:cov_iff_proof_comm_SINR} implies \eqref{eq:guar6} and \eqref{eq:cov_iff_proof_radar_SINR} implies \eqref{eq:guar7}, so \eqref{eq:guar} is feasible.

We now prove the `only if' direction. Assume \eqref{eq:guar} is feasible, and denote a feasible solution via $\widetilde{\bm{R}}$, $\{\widetilde{\bm{R_k}}\}_{k=1}^{K_c}$. Define the columns of $\bm{W_c}$ as
\begin{equation}
\bm{w_k} = \Big{(}\bm{h_k}^H\widetilde{\bm{R_k}}\bm{h_k}\Big{)}^{-1/2}\widetilde{\bm{R_k}}\bm{h_k},\quad k=1,\ldots,K_c.
\end{equation}
We note that for any $k=1,\ldots,K_c$,
\begin{equation}
\bm{w_k}\bm{w_k}^H = \frac{\widetilde{\bm{R_k}}\bm{h_k}\bm{h_k}^H\widetilde{\bm{R_k}}}{\bm{h_k}^H\widetilde{\bm{R_k}}\bm{h_k}},
\end{equation}
where we used the fact that $\widetilde{\bm{R_k}}$ is Hermitian. This means
\begin{equation}
\label{eq:comm_sig_power_app}
\bm{h_k}^H\bm{w_k}\bm{w_k}^H\bm{h_k} = \bm{h_k}^H\widetilde{\bm{R_k}}\bm{h_k}.
\end{equation}
Next, we prove the following lemma:
\begin{lemma}
\label{lemma:cov_iff}
For any $M\times M$ positive semidefinite matrix $\bm{A}$ and $M\times1$ vector $\bm{v}$ for which $\bm{v}^H\bm{A}\bm{v}\neq0$,
\begin{equation*}
\bm{B} \triangleq \bm{A} - \frac{\bm{A}\bm{v}\bm{v}^H\bm{A}}{\bm{v}^H\bm{A}\bm{v}} \in \mathcal{S}_M^{+}.
\end{equation*}
\end{lemma}
\begin{IEEEproof}
Consider any $M\times1$ vector $\bm{w}$. We have
\begin{equation}
\label{eq:cov_iff_lemma2}
\bm{w}^H\bm{B}\bm{w} = \bm{w}^H\bm{A}\bm{w} - (\bm{v}^H\bm{A}\bm{v})^{-1}\big{|}\bm{w}^H\bm{A}\bm{v}\big{|}^2.
\end{equation}
By the Cauchy-Schwarz inequality,
\begin{equation}
\label{eq:cov_iff_lemma3}
(\bm{v}^H\bm{A}\bm{v})(\bm{w}^H\bm{A}\bm{w}) \ge \big{|}\bm{w}^H\bm{A}\bm{v}\big{|}^2.
\end{equation}
Combining \eqref{eq:cov_iff_lemma2} and \eqref{eq:cov_iff_lemma3}, we see that $\bm{w}^H\bm{B}\bm{w} \ge0$. Since $\bm{w}$ is arbitrary, and $\bm{A}$ being Hermitian implies $\bm{B}$ is Hermitian, we conclude that $\bm{B} \in \mathcal{S}_M^{+}$, as desired.
\end{IEEEproof}
By Lemma~\ref{lemma:cov_iff}, we see that $\widetilde{\bm{R_k}}-\bm{w_k}\bm{w_k}^H\in\mathcal{S}_M^{+}$ for each $k$, which in turn means that
\begin{equation}
D \triangleq \sum_{k=1}^{K_c}\widetilde{\bm{R_k}} - \sum_{k=1}^{K_c}\bm{w_k}\bm{w_k}^H = \sum_{k=1}^{K_c}\widetilde{\bm{R_k}} -  \bm{W_c}\bm{W_c}^H \in \mathcal{S}_M^{+}.
\end{equation}
Since $\widetilde{\bm{R}}-\sum_{k=1}^{K_c}\widetilde{\bm{R_k}}\in\mathcal{S}_M^{+}$ by \eqref{eq:guar4} and $D\in\mathcal{S}_M^{+}$, we have
\begin{equation}
\widetilde{\bm{R}}-\sum_{k=1}^{K_c}\widetilde{\bm{R_k}} + D = \widetilde{\bm{R}}-\sum_{k=1}^{K_c}\bm{w_k}\bm{w_k}^H \in \mathcal{S}_M^{+}.
\end{equation}
Thus, we can choose $\bm{W_r}$ such that $\bm{W_r}\bm{W_r}^H = \widetilde{\bm{R}}-\bm{W_c}\bm{W_c}^H$ via eigendecomposition. This means that \eqref{eq:guar5} implies \eqref{eq:cov_iff_proof_power}.

Now we verify that \eqref{eq:cov_iff_proof_comm_SINR} is met. For any $k=1,\ldots,K_c$,
\begin{equation}
\xi_{oc}(\bm{W_c},\bm{W_r},k) = \xi_{oc}'(\widetilde{\bm{R}},\widetilde{\bm{R_k}},k) \ge \Gamma_c,
\end{equation}
where we used \eqref{eq:comm_sig_power_app} and the fact that $\widetilde{\bm{R}}$, $\{\widetilde{\bm{R_k}}\}_{k=1}^{K_c}$ is a feasible solution to \eqref{eq:guar}. The outer inequality means \eqref{eq:cov_iff_proof_comm_SINR} is satisfied.

Finally, we verify that the \eqref{eq:cov_iff_proof_radar_SINR} is met. For any $\theta\in\Theta$,
\begin{equation}
\xi_{or}(\bm{W_c},\bm{W_r},\theta,\xi_{ir}) = \frac{N_rM\xi_{ir}\bm{a}^H(\theta)\bm{W_r}\bm{W_r}^H\bm{a}(\theta)}{M\xi_{ir}\bm{a}^H(\theta)\bm{W_c}\bm{W_c}^H\bm{a}(\theta)+1}.
\end{equation}
Since $D$ is positive semidefinite, for any $\theta\in\Theta$, $\bm{a}^H(\theta)D\bm{a}(\theta)\triangleq d\ge0$. Since $N_r$, $M$, and $\xi_{ir}$ are positive, subtracting $N_rM\xi_{ir}d$ from the numerator and adding $M\xi_{ir}d$ to the denominator could not increase this fraction. Thus,
\begin{alignat}{1}
\label{eq:app_xi_or1}
\xi_{or}(\bm{W_c},\bm{W_r},\theta,\xi_{ir}) & \ge \frac{N_rM\xi_{ir}\bm{a}^H(\theta)(\widetilde{\bm{R}}-\sum_{k=1}^{K_c}\widetilde{\bm{R_k}})\bm{a}(\theta)}{M\xi_{ir}\bm{a}^H(\theta)(\sum_{k=1}^{K_c}\widetilde{\bm{R_k}})\bm{a}(\theta)+1} \\
\label{eq:app_xi_or2}
& = \xi_{or}'\bigg{(}\widetilde{\bm{R}},\sum_{k=1}^{K_c}\widetilde{\bm{R_k}},\theta,\xi_{ir}\bigg{)} \ge \Gamma_r,
\end{alignat}
where we used the fact that $\widetilde{\bm{R}}$, $\{\widetilde{\bm{R_k}}\}_{k=1}^{K_c}$ is a feasible solution to \eqref{eq:guar}, and \eqref{eq:cov_iff_proof_radar_SINR} is satisfied. This concludes the proof.

\section*{Acknowledgments}
\noindent Parts of this work were funded by MIT.

\bibliographystyle{IEEETran}
\bibliography{main}
\vfill

\end{document}